\def\be{\begin{equation}}
\def\te{\end{equation}}
\def\ba{\begin{eqnarray}}

\def\ta{\end{eqnarray}}

\def\l{\lambda}
\def\m{\mu}
\def\n{\nu}

\def\p{\pi}

\def\r{\rho}
\def\s{\sigma}

\newskip\humongous \humongous=0pt plus 1000pt minus 1000pt

\newif\ifdtup

\def\ha{{1\over 2}}
\def\la{\langle}
\def\ra{\rangle}

\def\(#1){(\ref{#1})}

\newcommand{\kl}[3]{\mbox{$\rm #1$}^{\mu\nu , \rho\sigma}_{#2}(#3)}

\documentstyle[prd,aps]{revtex}

\begin{document}
\draft
\title{Stochastic Gravity}
\author{B. L. Hu}
\address{Department of Physics,
         University of Maryland,
         College Park, MD 20742, USA.
         hub@physics.und.edu\\
-- Talk given at the Third Peyresq Meeting, France, June 1998.
To appear in Int. J. Theor. Phys.}
\date{Jan 30, 1999. Report umdpp \#99-062}
\maketitle
\begin{abstract}
We give a summary of the status of current research in stochastic semiclassical gravity and
suggest directions for further investigations.  This theory generalizes  the semiclassical
Einstein equation to an Einstein-Langevin equation with a stochastic source term arising 
from the fluctuations of the energy-momentum tensor of quantum fields.
We mention recent efforts in applying this theory to the study of black hole fluctuation
and backreaction problems, linear response of hot flat space, and structure
formation in inflationary cosmology.  To explore the physical meaning and implications 
of this stochastic regime in relation to both classical and quantum gravity,  we find it useful
to take the view that semiclassical gravity is mesoscopic physics and that general relativity is
the hydrodynamic limit of certain spacetime quantum substructures.    We view 
the classical spacetime depicted by general relativity as a collective state and the
metric or connection functions as collective variables.
Three basic issues - stochasticity, collectivity, correlations-  and
three processes - dissipation, fluctuations, decoherence- underscore the transformation from
quantum micro structure and interaction to the emergence of classical macro structure and
dynamics.   We discuss ways to probe into the high energy activity from below 
and make two suggestions: via effective field theory and  the correlation hierarchy.
We discuss how stochastic behavior at low energy in an effective
theory and how correlation noise associated with coarse-grained higher correlation functions in an
interacting quantum field could carry nontrivial information about the high energy sector.  Finally
we describe processes deemed important at the Planck scale, including tunneling and pair
creation, wave scattering in random geometry, growth of fluctuations and forms, Planck scale
resonance states, and spacetime foams.
\end{abstract}
\baselineskip=15pt
\pacs{PACS number(s):-04.62.+v, 05.40.+j, 98.80.Cq}


\section{Introduction}
\label{sec:intro}


\subsection{Semiclassical Gravity  from a Quantum Open System Viewpoint}

Starting from the well-cultivated and familiar terrain of quantum field
 theory in curved spacetime \cite{BirDav} in our search into deeper structures
  beyond semiclassical gravity with focus on the backreaction problem 
\cite{cpcbkr}, we came to a crossroad ten years ago. Having understood
the real physical meaning of dissipation in the effective dynamics of spacetime 
generated by the backreaction of particle creation \cite{CH87} with the help of
the Schwinger-Keldysh closed-time-path (CTP) formalism \cite{ctp},  we began to turn our
attention to possible existence of fluctuations generated by these processes. 
Following the dictums of nonequilibrium statistical mechanics, we proposed to 
view semiclassical gravity as a quantum open system \cite{Physica}.
The discrepancies which exist between the matter and gravity sectors (e.g.,
the heavy Planck mass which allows a Born-Oppenheimer approximation to be taken
in the transition of quantum cosmology to semiclassical gravity \cite{decQC})
enable one to treat classical spacetime as the `system'  of interest and
quantum matter field as the `environment' in the Langevin sense \cite{qos}.

This  then  prompted us to take a closer look at the influence functional 
approach (IF) \cite{if}  using  the quantum Brownian motion (QBM) as a model
\cite{qbm} because we were interested in a formalism which keeps the  
 self-consistency in treating  the  backreaction of the environment on the system, 
and displays the relation between dissipation, fluctuations, 
noise and decoherence \cite{Tsukuba}, the latter being a central issue in the 
investigation of the transition from quantum to classical \cite{envdec,conhis}.
Two sets of relations were of interest to us: The first set is between 
dissipation and fluctuations.  The second set is between noise and decoherence. 
 Fluctuation-dissipation relation is of course well-known 
\cite{FDR}, but it is usually assumed to be valid for systems at or close 
to equilibrium, and in fact usually derived with linear response theory 
\cite{LRT}. It would be easy to extract the fluctuations from the dissipation 
if such a relation holds also for ostensibly nonequilibrium systems like
in a cosmological backreaction problem, i.e., between classical dynamical
 spacetimes and evolving quantum fields. Whether such a relation exists in 
 semiclassical gravity is another crucial question asked in \cite{Physica}.   
If so,  what is the nature of such noises?
We posited that such a relation should also exist in  nonequilibrium systems  
such as that encountered in particle creation in a dynamical gravitational
  field. Our reasoning was that such a relation comes about as a relation 
  between two subsystems -- after one is being coarse-grained into an
   environment -- which when traced back should reflect the unitarity 
    condition for the dynamics of the original closed system.  With this, 
    we can then associate fluctuations or noise in the quantum field with  
 dissipation in the spacetime dynamics, and since dissipation is
generally nonlocal we asserted that the noise generated in particle creation
 would generally be multiplicative and colored.  These conjectures were
  realized in later investigations.

\subsection{Dissipation, Fluctuations, Noise and Decoherence}

In the same time frame when these questions about dissipation and noise
were investigated by the author,  the issue of decoherence of quantum systems
 and the emergence of classicality was pursued by a number of researchers coming from different
backgrounds in the 1980's using statistical mechanical concepts and methods. 
Specially relevant to our subject matter was
decoherence in quantum cosmology and the semiclassical gravity limit \cite{decQC}.
The source of noise and the role it plays  in this context was an important issue.
This brings in the second pair of relation mentioned above, that between noise and decoherence.
Such a relation was suggested in \cite{Tsukuba} using the Brownian motion model and Langevin
dynamics as a guide. Independently, Gell-Mann and Hartle \cite{GelHar2} in an excellent treatise
discussed how noise is instrumental to the emergence of classical equations of motion from
quantum dynamics and how it regulates the stability of classical structures.

On the technical level, the above evolution and linkage of concepts
on dissipation, noise and decoherence was facilitated by the closed-time-path,
influence functional and decoherence functional formalisms. Just as the 
Schwinger-Keldysh effective action \cite{ctp} enabled us  to get a real and causal equation of motion
\cite{CH87}, and the Feynman-Vernon influence functional \cite{if} enabled us to identify the noise kernel
 \cite{noisefield,HM2,CH94} and adopt the proper statistical mechanical interpretation 
of noise in quantum field theory, the decoherence functional (DF) of Gell-Mann and Hartle  
 and the consistent
 history formalism of Griffith and Omnes address the decoherence of histories and the
emergence of quasi-classical domains. These three formalisms are shown (or demonstrated in
specific models) to be closely related. (For the relation between CTP and IF, see \cite{CH94,Su},
that between IF and DF, see \cite{ifdf,dch,MarVer}). They constitute the formal basis for
establishing a  new regime between semiclassical and quantum physics named the 
stochastic (semiclassical) regime.

Thus viewing semiclassical gravity as an open system enabled us to
link up with inquiries of a fundamental nature such as the relation of classical, 
stochastic and quantum  and tap into the conceptual and technical resources in this
endeavor. It opened up a new horizon where some of the basic issues of quantum mechanics
 such as decoherence and the emergence of the classical world (classical
spacetime) can be addressed in statistical mechanical theoretical  terms; 
and the formal tools of quantum
field theory such as the effective action method can be used to  quantify statistical mechanical
notions and depict processes such as dissipation and noise (from activities of the quantum
vacuum of matter fields). With these ideas and methods at work, the stage was set around 1993
for probing into a deeper level of structure of gravity beyond the semiclassical theory, which we
called stochastic semiclassical gravity. (For a summary of work in this first stage 1989-1993, see,
e.g., \cite{Banff,HuSin})

\subsection{Einstein-Langevin Equation}

The next three years saw the developement of such a theory centering on
the quantification of noises associated with quantum field processes
\cite{noisefield} and the discovery of a new equation in this regime known as 
the Einstein-Langevin equation \cite{CH94,HM3,HuSin,CV96}
which relates the dissipative dynamics of spacetime and the fluctuations in the quantum matter 
fields. It has the form of a semiclassical Einstein equation (which contains a dissipative term from
the dissipation kernel in the influence action) but with an additional stochastic source term (from 
the noise kernel in the influence action).

Let me illustrate this theory with a brief sketch of the example of a conformally coupled 
scalar field in a weakly perturbed (anisotropic or inhomogeneous) spatially-flat
Friedmann-Lemaitre-Robertson-Walker  (FLRW)  Universe  with metric
 $ g_{\m\n}^{RW}$ plus small perturbations  $h_{\m\n}$,
\begin{equation}
    g_{\mu\nu}(x)= g_{\m\n}^{RW} +  h_{\m\n}
                       \equiv a(\eta)^{2} \tilde g_{\mu\nu}
   \label{eq:FLRW}
\end{equation}
Here $\eta$ is the conformal time related to the cosmic time $t$ by
$dt=a(\eta)d\eta$.
In this form  the metric  is conformally related (via conformal factor $a(\eta)$) to
the Minkowsky metric $\eta_{\mu\nu}$ and its perturbations $\tilde h_{\mu\nu}(x)$:
\be
\tilde g_{\m\n} =\eta_{\mu\nu}+\tilde h_{\mu\nu}(x)
\te
The perturbations $ h_{\m\n}$ can be homogeneous (the case $=\delta a^2$ 
was treated by Calzetta and Hu  \cite{CH94}),
or  anisotropic (as in a Bianchi Type I case treated by Hu and Sinha \cite{HuSin}), 
or inhomogeneous (treated  by Campos, Martin and Verdaguer \cite{CV96,MarVer}). 
 Here we follow the latter work.

The classical action for a free massless real
scalar field $\Phi(x)$ is given by
\begin{equation}
   S_f[g_{\mu\nu},\Phi] =
        -{1\over2}\int d^nx\sqrt{- g}
            \left[g^{\mu\nu}\partial_\mu\Phi\partial_\nu\Phi
                 +\xi(n) R\Phi^2
            \right]
\end{equation}
where $R$ is the Ricci curvature scalar for the metric $ g_{\mu\nu}$
and  $\xi(n)=(n-2)/[4(n-1)]$, n being the spacetime dimension,  is the 
coupling of the field to the spacetime, with $\xi(4) = 0, 1/6$ corresponding
to minimal and conformal couplings respectively.
We consider a massless conformally coupled scalar field here.
 Define a conformally related field
$\tilde \Phi(x)\equiv a(\eta)^{(n/2-1)} \Phi(x)$,
the action $S_f$ (after one integration by parts) 
\be
S_f[\tilde g_{\mu\nu},\tilde \Phi]=
        -{1\over2}\int d^nx\sqrt{-\tilde g}
            \left[\tilde g^{\mu\nu}\partial_\mu\tilde \Phi\partial_\nu\tilde \Phi
                 +\xi(n)\tilde R\tilde \Phi^2
            \right],
\te
takes the form of an action for a free massless conformally coupled real
scalar field $\tilde \Phi(x)$ in a spacetime with metric $\tilde g_{\mu\nu}$ --
In this case it is a  nearly flat spacetime. As the physical field $ \Phi(x)$
is related to the field $\tilde \Phi(x)$ by a power of the conformal factor
a positive frequency mode of the field $\tilde \Phi(x)$ in flat spacetime will
correspond to a positive frequency mode in the conformally related
space. One can thus establish a quantum field theory in the conformally
related space by use of the conformal vacuum (see \cite{BirDav}).
Quantum effects such as particle creation arises from
the breaking of conformal flatness of the spacetime
produced by the perturbations $ h_{\mu\nu}(x)$.

\subsubsection{Semiclassical Einstein Equation}

The Einstein equation for classical gravity is
\be
G_{\mu\nu}[g] + \Lambda g_{\mu \nu}
        =8\pi G  T_{\mu\nu}^{c}
\te
where
G is the Newton constant,  $\Lambda$ is the cosmological constant, 
$g_{\mu\nu}$ is the spacetime metric, $G_{\mu\nu}$ is the Einstein tensor,
and $T_{\mu\nu}^c$ is the energy momentum tensor of  classical matter or fields.
 One now adds the quantum field as a source, 
and gets the semiclassical Einstein equation (SCE)
\begin{equation}
    G_{\mu\nu}[g] + \Lambda g_{\mu \nu}
        =8\pi G  \left( T_{\mu\nu}^{c} +  T_{\mu\nu}^{q} \right)
\label{SCE}
\end{equation}
where  $T_{\mu\nu}^q \equiv \langle T_{\mu\nu} \rangle_q$ 
is the expectation value of the stress tensor operator in some quantum state
of the matter field $\Phi$.
In general there are ultraviolet divergences in  $\langle T^{\mu\nu} \rangle_q$. To 
remove or cure them one introduces regularization or renormalization procedures
by adding counter terms or absorbing them into the cosmological constant,
the Newton constant and the coupling constants of the curvature-squared terms 
corresponding to the quartic, quadratic and logarithmic divergences \cite{BirDav}. 
As a result the renormalized  SCE equation takes the form 
\begin{equation}
\left( G_{\mu\nu}[g]+\Lambda g_{\mu\nu}\right)
-l_P^2 \left(\alpha A_{\mu\nu} +\beta B_{\mu\nu}\right)[g]=
{8\pi G}\langle \hat T^R_{\mu\nu}\rangle [g],
\label{D1}
\end{equation}
where $G$, $\Lambda$, $\alpha$ and $\beta$ are now renormalized coupling constants
and $l_P \equiv \sqrt {16\pi G}$ is the Planck length.
$A_{\mu\nu}$ and $B_{\mu\nu}$ are divergenceless local curvature tensors defined
by 
\begin{eqnarray}
     A^{\mu\nu}(x)
&\equiv&  \frac{1}{\sqrt{-g}}\frac{\delta}{\delta g_{\mu\nu}}
\int d^4x \sqrt{-g} C_{\alpha\beta\rho\sigma}C^{\alpha\beta\rho\sigma}
\nonumber \\
      &=&{1\over2}g^{\mu\nu}C_{\alpha\beta\rho\sigma}
                                  C^{\alpha\beta\rho\sigma}
               -2R^{\mu\alpha\beta\rho}{R^\nu}_{\alpha\beta\rho}
               +4R^{\mu\alpha}{R_\alpha}^\nu
         \nonumber \\
      & &\hskip .1cm
               -{2\over3}RR^{\mu\nu}
               -2\Box_g R^{\mu\nu}
               +{2\over3}R^{;\mu\nu}
               +{1\over3}g^{\mu\nu}\Box_g R,
 \end{eqnarray}
where $C_{\alpha\beta\rho\sigma}$ is the Weyl tensor, and
\begin{eqnarray}
B^{\mu\nu}(x)
      &\equiv& \frac{1}{\sqrt{-g}}\frac{\delta}{\delta g_{\mu\nu}}
\int d^4x \sqrt{-g} R^2   \nonumber\\
       & = & {1\over2}g^{\mu\nu}R^2
               -2RR^{\mu\nu}
               +2R^{;\mu\nu}
               -2g^{\mu\nu}\Box_g R,
\end{eqnarray}
where $R$ is the Ricci scalar.   The divergence-free tensor
$\langle \hat T^R_{\mu\nu}\rangle [g]$ is the expectation value in some
quantum state of the renormalized stress tensor operator
$\hat T^R_{\mu\nu}[g]$ of the field $\hat\Phi$. 

For a massless conformally coupled
scalar field in the metric  (1) above, $\langle T^{\mu\nu} \rangle_q$
has the form ( the subscripts 0, 1 in parentheses denote zeroth and first order in 
$h_{\mu\nu}$) \cite{CV96}
\begin{eqnarray}
   \langle T^{\mu\nu}_{(0)} \rangle_q
        &=& \l\left[   H^{\mu\nu}_{(0)}
                        -{1\over 6}   B^{\mu\nu}_{(0)}
                  \right]
           \nonumber \\
   \langle T^{\mu\nu}_{(1)} \rangle_q
        &=& \l\left[ \atop \right.
                         (  H^{\mu\nu}_{(1)}
                          -2  R^{(0)}_{\alpha\beta}
                            C^{\mu\alpha\nu\beta}_{(1)}
                         )
                        -{1\over 6}   B^{\mu\nu}_{(1)}
              \nonumber \\
          & &\hskip1cm
                        +3a^{-3}
                            \left( -4( C^{\mu\alpha\nu\beta}_{(1)}
                                       ln a )_{,\alpha\beta}
                        +\int d^4y A^{\mu\nu}_{(1)}(y)
                                      \mbox{\rm K}(x-y;\bar\mu)
                           \right)
                  \left. \atop \right].
   \label{eq:quantum stress tensor}
\end{eqnarray}
where the constant $\l = 1/2880\p^2$ characterizes one-loop quantum correction
terms (which include the trace anomaly and particle creation processes) and $\bar \mu$
 is a renormalization parameter.
Here  $H^{\mu\nu}(x)$ arises from the counterterms in the renormalization 
of the energy mementum tensor (see, e.g.,  \cite{FulPar}) and is related to
 $A^{\mu\nu}, B^{\mu\nu}$ above:
\be
 H^{\mu\nu}(x)
     \equiv -R^{\mu\alpha}{R_\alpha}^\nu
               +{2\over3}RR^{\mu\nu}
               +{1\over2}g^{\mu\nu}R_{\alpha\beta}R^{\alpha\beta}
               -{1\over4}g^{\mu\nu}R^2.
\te
I call attention to the exisitence of a dissipation  term (kernel $K$) above
describing the backreaction of particle creation on the background spacetime
dynamics \cite{CH87,CV94}.
All terms  in the semiclassical Einstein equation originating from renormalization
and backreaction, including the dissipative kernel, are familiar from model calculations
done in the Seventies and Eighties (see. e.g., \cite{cpcbkr}). One needs the CTP 
effective action \cite{ctp} to derive the correct SCE which is real and causal
 \cite{CH87,CV94}.

\subsubsection{Stochastic Semiclassical Einstein Equation}

The stochastic semiclassical Einstein, or Einstein-Langevin  equation  (ELE)
\cite{CH94,HM3,HuSin} differs from the 
semiclassical Einstein equation (SCE) by the presence of a stochastic term
 measuring the fluctuations of quantum sources
(arising from the difference of particles created in neighboring histories,
see, \cite{CH94}) which is intrinsically linked to the dissipation term in 
the dynamics of spacetime.
Two points are noteworthy: a) The fluctuations and dissipation kernels
(decipherable from the influence action)
obey a fluctuation -dissipation relation, which embodies the backreaction
effects of quantum fields on classical spacetime.
b) The stochastic source term engenders metric fluctuations.

The semiclassical Einstein equation depicts
a mean field theory which one can  retrieve from the Einstein-Langevin equation by
taking a statistical average with respect to the noise distribution.
We used the influence functional formalism to extract these new information
\cite{HuSin,CV96}.
The stochastic semiclassical Einstein equation, or Einstein-Langevin equation, 
takes on the form
\begin{eqnarray}
     G_{\mu\nu}[g] + \Lambda g_{\mu\nu}
        &=& 8\pi G ( {T_{\mu\nu}}^c +  {T_{\mu\nu}}^{qs})
           \nonumber \\
        T_{\mu\nu}^{qs}
        &\equiv& \langle T_{\mu\nu} \rangle_{q} +  T_{\mu\nu}^{s}
   \label{ELE}
\end{eqnarray}
The new term $T_{\mu\nu}^{s}= 2 \tau_{\mu\nu}$ which is of classical  
stochastic nature,  measures the  fluctuations of the energy momentum tensor 
of the quantum field. Define
\be
{\hat t}_{\mu\nu}(x) \equiv {\hat T}_{\mu\nu}(x) - \la {\hat T}_{\mu\nu}(x)\ra \hat I
\te
(Such a tensor is computed in the backgound metric,  not  the perturbed metric.)
It is related to the noise kernel $N_{\mu\nu\rho\sigma}$ bitensor by
\be
4 N_{\mu\nu\rho\sigma} (x,y) \equiv \ha \la \{ {\hat t}_{\mu\nu}(x), 
{\hat t}_{\rho\sigma}(y) \} \ra 
\label{D4}
\te
where $\{ \}$ means taking the symmetric product. The noise kernel appears
in the real part of the influence action.

The noise kernel is free of ultraviolet divergence as one can see from its
definition and the fact that the ultraviolet behavior of 
$\hat T_{\mu\nu}$ and $\langle \hat T_{\mu\nu}\rangle$ is the same; thus one
can replace $\hat T_{\mu\nu}$ by $\hat T^R_{\mu\nu}$ in these equations. 
The noise kernel defines a real classical Gaussian stochastic symmetric tensor field
$\tau_{\mu\nu}$ which is characterized to lowest order by the following correlators,
\begin{equation}
\langle\tau_{\mu\nu}(x)\rangle_\tau=0,\ \ \ \ 
\langle \tau_{\mu\nu}(x) \tau_{\rho\sigma}(y)\rangle_\tau= 
N_{\mu\nu\rho\sigma}(x,y),
\label{D6}
\end{equation}
where $\langle\,\rangle_\tau$ means taking a statistical average  (for simplicity no higher
order correlations are assumed). Since $\hat T^R_{\mu\nu}$ is
self-adjoint one can see that $N_{\mu\nu\rho\sigma}$ is  symmeric, real,
positive and semi-definite.  Furthermore, as a consequence of (\ref{D4})
and the consevation law $\nabla^\mu \hat T^R_{\mu\nu}=0$, this stochastic tensor
is divergenceless in the sense that $\nabla^\mu
\tau_{\mu\nu}=0$ is a deterministic zero field. Also 
 $g^{\mu\nu}{\tau}_{\mu\nu}(x) = 0$,  signifying that there is no  
stochastic correction to the trace anomaly (if $T_{\mu\nu}$ is traceless). 
Here  all covariant
derivatives are taken with respect to the background metric $g_{\mu\nu}$ which
is a solution of the semiclassical equations. Taking the statistical average of 
 (\ref{ELE} ), as a consequence of the noise correlation relation (\ref{D6}),
\begin{equation}
          \langle T_{\mu\nu}^{qs}
          \rangle_\xi
        = \langle T_{\mu\nu} \rangle_q
\end{equation}
we recover the semiclassical Einstein equation (\ref{SCE}).
%

Now for a spacetime with background metric $g_{\mu\nu}$ and weak gravitational
perturbation $h_{\mu\nu}$ the EL equation to linear order in $h_{\mu\nu}$
has the form
\begin{equation}
\left( G_{\mu\nu}[g+h]+\Lambda (g_{\mu\nu}+h_{\mu\nu})\right)
-2 l_P^2 \left(\alpha A_{\mu\nu} +\beta B_{\mu\nu}\right)[g+h]=
8\pi G (\langle \hat T^R_{\mu\nu}\rangle [g+h]+2\tau_{\mu\nu}),
\label{D7}
\end{equation}
 The symmetry and
divergenceless of the stochastic tensor in the background metric guarantee the
consistency of this semiclassical Einstein-Langevin equation. This equation
gives the first order correction to semiclassical gravity in the sense that it
incorporates the correlation of $T_{\mu\nu}$. The distinct feature is that it
predicts the existence of a stochastic component in the metric which we call  
metric {\it fluctuations}. It is induced by the quantum stress tensor fluctuations.
 Since the stress tensor fluctuations are defined on the
background metric $g_{\mu\nu}$, the stochastic field $\tau_{\mu\nu}$ 
does not depend on the metric {\it perturbations}  $h_{\mu\nu}$. 
Therefore Eq. (\ref{D7}) is
a linear stochastic equation for $h_{\mu\nu}$ with an inhomogeneous term
$\tau_{\mu\nu}$, its solution can be formally written as the functional
$h_{\mu\nu}[\tau]$.
Taking the statistical
average of Eq. (\ref{D7}) one sees that the metric $g_{\mu\nu}+
\langle h_{\mu\nu}\rangle_\xi$ must be a solution of the semiclassical Einstein
equation linearized around $g_{\mu\nu}$. By the gauge invariance of Eq.
(\ref{D7}) it is clear that if $h_{\mu\nu}$ is a solution of this equation,
$h_{\mu\nu}'=h_{\mu\nu}+\nabla_\mu\zeta_\nu  +\nabla_\nu\zeta_\mu $, where
$\zeta_\mu(x)$ is a Gaussian stochastic field on the background spacetime, is
a physically equivalent solution.


For the example of a perturbed spatially flat FLRW universe with a quantum scalar field 
$\Phi$ the tensor $\tau_{\mu\nu}(x)$ is given by \cite{CV96}
\begin{equation}
   \tau_{\mu\nu}(x)=
       -2\partial_\alpha\partial_\beta\xi^{\mu\alpha\nu\beta}(x),
   \label{eq:source}
\end{equation}
which is seen to be symmetric and traceless, {\it i.e.}
$\tau_{\mu\nu}(x)=\tau_{\nu\mu}(x)$ and $\tau^\mu_{\,\,\mu}(x)=0$.
The stochastic correction to the stress tensor has vanishing divergence with respect to 
the background metric.

 In this problem  the tensor
$\xi_{\mu\nu\alpha\beta}(x)$ has the symmetries
of the Weyl tensor, {\it i.e.} it has the symmetries of the Riemann tensor 
and vanishing trace in all its indices.
It is characterized completely by the noise kernel $N(x-y)$
(the probability distribution for the noise is Gaussian) \cite{HuSin,CV96}
\begin{eqnarray}
      \langle\xi_{\mu\nu\alpha\beta}(x)\rangle_{\xi}
  &=& 0,
      \nonumber \\
      \langle\xi_{\mu\nu\alpha\beta}(x)
      \xi_{\rho\sigma\lambda\theta}(y)\rangle_{\xi}
  &=& T_{\mu\nu\alpha\beta\rho\sigma\lambda\theta}
      \mbox{\rm N}(x-y),
   \label{eq:gaussian correlations}
\end{eqnarray}
Here $T_{\mu\nu\alpha\beta\rho\sigma\lambda\theta}$ is
the product of four metric tensors (in such
a combination that the right-hand side of the equation satisfies
the Weyl symmetries of the two stochastic fields on the left-hand side).
Its explicit form is given in \cite{CV96}
 
As mentioned above the new source term $2\tau_{\mu\nu}$ 
will produce a stochastic contribution $h^{s}_{\mu\nu}$ to the spacetime metric,
{\it i.e.} $h_{\mu\nu}=h^c_{\mu\nu}+h^{s}_{\mu\nu}$. 
Considering a flat background spacetime (setting $a=1$ in (\ref{eq:FLRW}) 
and dropping the tilde on $h_{\mu\nu}$ for simplicity), 
one obtains, by adopting the harmonic gauge
condition $(  h^{s}_{\mu\nu} - {1\over2}\eta_{\mu\nu}  h^{s})^{,\nu}=0$,
a linear equation for the metric fluctuations (off Minkowski spacetime here) $  h^{s}_{\mu\nu}$
\begin{eqnarray}
   \Box   h^s_{\mu\nu}
        &=& 16\pi G  T^s_{\mu\nu},
           \nonumber \\
   T_{\mu\nu}^s
        &=& 2   \tau_{\mu\nu}
         =-4\partial_\alpha\partial_\beta   \xi^{\mu\alpha\nu\beta},
\end{eqnarray}
 The computation of the noise correlations and the solution of these equations 
have been given by Campos and Verdaguer \cite{CV96}.   
 Calzetta, Campos and Verdaguer have solved the Einstein-Langevin 
equation for a cosmological problem with both noise and fluctuations \cite{CCV}.
\footnote{ A comment from E. Verdaguer: 
In equation (20)  Campos and Verdaguer neglected the dissipation term, or more 
       precisely the expectation value of the  stress tensor to
      linear order in $h_{\mu\nu}$ which is of the same order as
      $S_{\mu\nu}$ is stochastic. For this reason the solution they found
      was only formal, it is divergent in fact. One can get a finite
      result if one starts the perturbation at some initial time zero
      (in the paper they start at $t=-\infty$). This is similar to having
      a particle in a bath with no dissipation for a very long time,
      the fluctuations will take it very far. In \cite{CCV} the  calculation was
      only for the homogeneous conformal mode. }
Recently        Martin and Verdaguer  \cite{MarVer} have revisited this problem.
They  solved the stochastic semiclassical Einstein equation Eq. (\ref{D7})  around the
 Minkowski spacetime $\eta_{\mu\nu}$
for a massless conformally coupled scalar field in its vacuum state
$|0\rangle$. In this case $\langle 0|T^R_{\mu\nu}[\eta]|0\rangle=0$
and if we take $\Lambda=0$, the Minkowski metric is a  trivial
solution of the semiclassical Einstein equation  (\ref{D1}).
Since the vacuum state is not an eigenstate of
$\hat T^R_{\mu\nu}[\eta]$ fluctuations of the stress tensor are present. Eq.
(\ref{D7}) reduces in this case to the linearized SCE equations derived by 
Horowitz \cite{Hor} for studying 
the semiclassical stability of flat spacetime, but with a new inhomogeneous source
term $\tau_{\mu\nu}$. Martin and Verdaguer evaluated the two point correlation function
of the linearized Einstein tensor and  found that for
spacelike separated points $\vec x$ and $\vec x'$ it goes like 
\be
\frac{1}{l_P^2}\frac{1}{|\vec x-\vec x'|^2}\exp\left(-\frac{|\vec x-\vec
x'|}{l_P}\right), \nonumber 
\te
The above result shows that the quantum field
fluctuations induce metric fluctuations with a correlation length $l_P$.
The appearence of Planck length here is not
surprising since for a massless scalar field coupled to gravity there is no
other length scale in the problem. It is noteworthy that this result is
not analytic in $l_P$ and thus it could not have been obtained by a
perturbative expansion in the Planck length. Of course, this
semiclassical result is expected to break down at Planck scale and 
quantum fluctuations of the metric beyond that induced by linear
perturbations (gravitons can be treated as quantum field source as each
is identical to two components of massless minimally coupled scalar field) 
would become important. 

For other  recent developments, I would like to mention a derivation of the 
E-L equation  from renormalization group considerations by Lombardo 
and Mazzitelli \cite{LomMaz},
and the application of the CTP-IF formalism to the study of backreaction of
Hawking radiation in  2D dilatonic black hole spacetimes by Lombardo, Mazzitelli
and Russo \cite{LMR}.

\subsection{Stochastic in relation to Semiclassical and Quantum Gravity}

Stochastic gravity is a regime intermediate between semiclassical and quantum 
gravity. It is perhaps instructive to examine the distinction among these three 
theories  (this was  displayed in one slide in  Dr. Verdaguer's lecture).

We use the example above for gravitational perturbations $h_{\mu\nu}$ in a 
FLRW universe with background metric $g_{\mu\nu}$ driven by the expectation
value of the energy momentum tensor of a scalar field $\Phi$, as well as its fluctuations
 ${\hat t}_{\mu\nu}(x)$.
Let us compare the stochastic with the semiclassical and quantum equations 
of motion for the metric perturbation field $h$ (we will use schematic notations for simplicity).
The semiclassical equation is given by
\be
\Box h = \la \hat T \ra
\te
where $\la \ra$ denotes taking the quantum average (e.g., the vacuum expectation value) 
of the operator enclosed. Its solution can be written in the form
\be
h = \int G \la \hat T\ra, ~~~~~
h_1h_2 = \int\int G_1G_2 \la \hat T\ra \la\hat T\ra.
\te
The quantum (Heisenberg) equation
\be
\Box \hat h = \hat T
\te
has solutions
\be
\hat h = \int G \hat T, ~~~~~
\la \hat h_1 \hat h_2\ra = \int \int G_1 G_2 \la \hat T \hat T \ra_{\hat h, \hat \phi}
\te
where the average is taken  with respect to
the quantum fluctuations of both the gravitational and the matter fields.
Now for the stochastic equation, we have
\be
\Box h = \la \hat T \ra + \tau
\te
with solutions
\footnote{In this schematic form we have not displayed the homogeneous
 solution carrying the information of the (maybe random) initial condition. 
This solution will  exist in general, and may even be dominant if dissipation is weak .
When both the uncertainty in initial conditions and
the stochastic noise are taken into account, the Einstein - Langevin
formalism reproduces the exact graviton two point function, in the
linearized approximation. Of course, it fails to reproduce the expectation
value of observables which could not be written in terms of graviton
occupation numbers, and in this sense it falls short of full quantum gravity. 
I thank  E. Calzetta for this comment.}

\be
h = \int G \la \hat T \ra + \int G \tau, ~~~~~
h_1 h_2 = \int\int G_1 G_2 [ \la \hat T \ra \la \hat T \ra + 
(\la \hat T \ra \tau + \tau \la \hat T \ra) + \tau \tau]
\te
We now take the noise average $\la \ra_\xi$ . Recall that the noise
is defined in terms of the stochastic sources $\tau$ as
\be
\la \tau \ra_\xi = 0, ~~~~ \la \tau_1\tau_2\ra_\xi \equiv \la \hat T_1 \hat T_2 \ra - \la \hat T_1\ra
\la \hat T_2\ra
\te
we get
\be
\la h_1 h_2 \ra_\xi = \int \int G_1 G_2 \la \hat T \hat T \ra_{\hat \phi}
\te
Note that the correlation of the energy momentum tensor appears just like in the quantum case,
but the average here is only over noise from quantum fluctuations of the matter field. 

As seen above, while the semiclassical regime describes the effect of a quantum  matter field only 
through its mean value (vacuum expectation value), the stochastic regime includes the fluctuations
of quantum fields as reflected in the new stochastic term in the energy momentum tensor. Thus
stochastic gravity carries some information about the correlation of fields 
(and the related phase information)
which is absent in semiclassical gravity. Here we have invoked the relation between fluctuations
and correlation, a variant form of the fluctuation-dissipation relation. This feature pushes
stochastic gravity  closer than semiclassical gravity to quantum gravity in that the correlation
in quantum field and geometry fully present in quantum gravity is partially retained in stochastic
gravity, and the background geometry has a way to sense the correlation of the quantum fields
through the noise term in the Einstein-Langevin equation, which shows up as metric
fluctuations.

Thus `noise' as used in this more precise  language and context is not something one 
can arbitrarily assign  or relegate, but has taken on 
a wider meaning in that it embodies the contributions of the higher correlation functions in the
quantum field. Only the lowest order is being displayed in what have been done so far, in terms of
the 2 point function of the energy momentum tensor (or the 4 point function of fields). Although
the Feynman- Vernon way can only accomodate Gaussian noise of the matter fields and takes a
simple form for linear coupling to the background spacetime, the notion of noise can be made
more general and precise. (For an example of more complex noise associated with more
involved backreactions arising from strong or nonlocal coupling, see Johnson and Hu
\cite{JohHu}). Progress is made now on how to characterize the higher order correlation
functions of an interacting field systematically from the Schwinger-Dyson equations in terms of
what Calzetta and I called `correlation noise' \cite{cddn,StoBol}, after the BBGKY hierarchy.
This will be discussed in a later section.  

Notice also that the difference between stochastic gravity and quantum gravity is that in the
former only the fluctuations and correlations of matter fields are accounted for while the full
quantum theory should also include the fluctuations and correlations  of the quantum gravitational
field. We will focus on this difference and discuss how closely one could probe into the full theory 
with stochastic equations later.

The aim of this paper is to deliberate on the meanings of this new regime, the significance of
quantum noise and metric fluctuations in affecting Planck scale processes and how correlation
bears to reveal a deeper level of spacetime structure short of knowing the full theory of quantum
gravity. I will also describe some ongoing research in this program and  make suggestions for
further investigations.

\section{Metric Fluctuations from Backreaction of Quantum Fields}

By construction this new framework is suitable for investigation into metric fluctuations and
backreaction effects. So far it has been applied  \cite{HuSin,CV96,CCV} to study quantum
effects in cosmological spacetimes. Work on black hole spacetimes has just begun
\cite{Vishu,Phillips,HPR,HRS}.  Parallel to this is the interesting application
 to noise-induced phase transitions which is described in Dr. Calzetta's talk  \cite{CalVer}.
 I will also mention other
directions, including applications in thermal field theories (hot flat space) \cite{CamHu}.

\subsection{Metric Fluctuations in Semiclassical Gravity}

Metric fluctuation and its more colorful generalization called spacetime foam have been a subject
of intermittant speculations since Wheeler introduced it in the early 60's to address `the issue of the final
state' in general relativity  \cite{Wheeler}. We will have more to say about this generalization
from the viewpoint of stochastic gravity later. Here it is sufficient to point out that the correlation
functions for the noise kernels obtained by Calzetta, Hu, Matacz and Sinha \cite{CH94,HM3,HuSin} and 
Campos, Martin and Verdaguer \cite{CV96,MarVer} (see last section) give the first quantitative 
description of metric fluctuations as induced by quantum fields.
To begin, it is perhaps useful to emphasize the difference in the meaning of `metric fluctuations'
used in our program which includes backreaction from quantum fields and that  used by many 
others in the test field context, where one considers classical gravitational perturbations
$h_{\m\n}$  from a fixed background geometry
and their two-point functions $\langle h_{\m\n}(x) h_{\r\s}(y)\rangle$ (averaged
with respect to some vacuum, in a semiclassical sense).
 \footnote{The two point function of  gravitons are not stochastic variables and so in a stricter
sense they should not be called  metric `fluctuations'. To avoid confusion we may at times call 
our quantities $h^{s}_{\m\n}$ induced metric fluctuations.}
It is useful as a measure of the fluctuations in the gravitational field at particular regions of
spacetime. Ford and coworkers have explored this aspect in great detail \cite{Ford}
 (They call this kind of fluctuations `active' and the kind we
discuss here `passive'-- I would prefer to call them `spontaneous' and `induced'.)  However, when
backreaction is included, as is necessary at the Planck scale, with the background spactime metric
 and the quantum fields present evolving together consistently, the graviton 2-point function 
calculated with respect to
a fixed background (as in the case of `active' fluctuations) rapidly loses its relevance.

In contrast, metric fluctuations $h^{s}_{\m\n}$ here \cite{CH94,HM3,HuSin,CV96} are
defined for semiclassical gravity in the backreaction context.  They are classical stochastic
quantities arising from the flucutations in the quantum fields present and are important only at the
Planck scale.   We see that they are derived from the noise kernel, which,  if the quantum field is
 the graviton,  involves graviton 4-point
functions. It is this quantity which enters into the fluctutation-dissipation relation -- not the usual
graviton 2 point function -- which encapsulates the semiclassical backreaction. 

An immediate application of
metric fluctuations is on the stability of semiclassical spacetimes (solutions to the semiclassical
Einstein equations) against stochastic sources from particle creations, and the validity of
semiclassical gravity.  The determining factor is in the noise kernel, which is related to the
fluctuations of the energy momentum tensor. Kuo and Ford \cite{KuoFor} have calculated the
fluctuations in the Casimir energy density for flat space and found it to be comparable to the
mean. Phillips and Hu \cite{PhiHu} confirmed their result using a covariant generalized 
zeta function method. 


For quantum fields in a curved spacetime with an Euclidean section, Phillips and Hu \cite{PhiHu}
derived a general expression for the stress energy tensor two-point function in terms of the
effective action.  The renormalized two-point function is given in terms of the second variation of
the Mellin transform of the trace of the heat kernel for the quantum fields. For systems in which
a spectral decomposition of the wave opearator is possible,  one can derive an exact expression for 
this two-point function. As a measure of the magnitude of fluctuations, 
they used the dimensionless expression \cite{KuoFor}
 for the ratio between the variance
of each component of the stress-energy tensor compared  to the mean:
\be
\Delta_{abcd}(x) = \left|
  \frac{ \left< T_{ab}(x) T_{cd}(x) \right>_{\rm ren} - 
     \left< T_{ab}(x)\right>_{\rm ren} \left< T_{cd}(x) \right>_{\rm ren}}
       { \left< T_{ab}(x) T_{cd}(x) \right>_{\rm ren} }
                \right|
\te
From inspection, $0 \le \Delta_{abcd} \le 1$. Only for $\Delta \ll 1$
can the fluctuations be viewed as small. On the other hand, $\Delta \sim 1$
indicates that the fluctuations can be large compared to
the mean value. Phillilps and Hu studied two cases in details with this method:
$d$ dimensional flat space product a circle ($R^d\times S^1$) with a 
minimally coupled massless scalar field,  and the Einstein universe ($S^3$) with a 
conformally coupled massless scalar field.
 The results for the energy density are ($\tau$ denotes Euclildean time):
\be
\Delta_{\tau\tau\tau\tau}\left( R^d\times S^1\right) = 
   \frac{(d+1)(d+2)}{(d+1)(d+2)+2}
\quad {\rm and} \quad
\Delta_{\tau\tau\tau\tau}\left( S^3 \right) = \frac{111}{112} \sim .99
\te
 The large variance signifies the importance of quantum fluctuations and
may indicate the breakdown of semiclassical gravity at sub-Planckian scales.

\subsection{Black Hole Fluctuations and Backreaction}

Work in progress now focuses on fluctuations of the energy density of quantum fields in early
universe and black hole spacetimes. These results will have direct bearings on structure formation from
quantum fluctuations in the early universe (see, e.g. \cite{CH95} and references therein) and 
 stability of black holes against
Hawking radiation and the related entropy and information loss issues. 
Here, as before, the central task is the computation of the noise kernel, or the fluctuations 
of the energy momentum tensor.
One can use the zeta function method (as in Phillips and Hu \cite{PhiHu}) for treating the second
variation of the effective action, or more explicitly, the covariant point splitting method
\cite{ChrDeW}. The main difficulty for black hole spacetimes, as is already present in the
calculation of the  regularized energy momentum tensor for spherically symmetric spacetimes
\cite{AndHis}, lies in the radial functions. For optical metrics one can use the Gaussian
approximation for the propagators as was done by Page \cite{Page}, who obtained an expression
for the energy density of quantum scalar fields which was shown to be good to an unexpectedly
high accuracy. Phillips in his thesis \cite{Phillips} has obtained results for the fluctuations of the
energy density of a scalar field in a general optical metric and is in progress for the Schwarzschild
metric at the horizon.   Earlier, Ford \cite{Ford97} has shown that (spontaneous) black hole
horizon fluctuations -- the graviton two point function -- are much smaller than Planck dimensions
for black holes whose mass exceeds the Planck mass. From our result and Ford's one sees that, 
contrary to some recent claims \cite{CEIMP}, the semiclassical  derivation  of Hawking radiance
should remain valid  for black holes larger than the Planck mass and there is no 
drastic effect near the horizon arising from metric fluctuations. 
Other recent work on black hole
horizon (spontaneous) fluctuations include Sorkin and Frolov et al  \cite{Sorkin,BFP}.

The cosmological backreaction problem saw two stages of development  as represented by  the
use of  the in-out cum in-in effective action (e.g. \cite{CH87}) followed by the influence action
(e.g., \cite{HuSin}) for extracting first the mean value and then the fluctuations of the energy
momentum tensor, which physically corresponds to the study of dissipation and fluctuations of the
spacetime. Likewise, black hole backreaction problem also progressed in  two stages. The first
stage started in the early 80's with the work of Candelas, Howard, Page, Frolov, Jensen,
McLaughlin, Ottiwell, Hiscock, Anderson and others (see \cite{AndHis} and references therein) in the
calculation of the regularized energy momentum tensor for quantum fields in  black hole
spacetimes. The second stage has just begun. It focuses on  calculating the fluctuations of the
energy momentum tensor as described above, and with it the backreaction on the black hole
spacetime configurations and dynamics. In \cite{HRS} Raval, Sinha and I have given a loose
sketch of our program of investigation. We discussed the formulation of the problem, commenting
on possible advantages and shortcomings of existing works,  and introduced our own approach
via  stochastic semiclassical theory of gravity. The goal is to derive and solve the
Einstein-Langevin equation (or its physical equivalent, the fluctuation-dissipation relation) for a
self-consistent description of metric  fluctuations and the dissipative dynamics of a black hole with
backreaction from its radiance. We have divided the problem into two main classes, the
quasi-static problem and the dynamic problem. The quasi-static problem is  characterized by a
black hole in quasi-equilibrium with its Hawking radiation (enclosed in a box to ensure relative
stability).  One important early work on backreaction of this kind is by York \cite{York}, while
the most thorough to date is carried out by Hiscock, Anderson et al \cite{AndHis}.  Backreaction
for dynamical (collapsing) black holes are much more difficult  to treat than static ones, and there
are fewer viable attempts.  For situations with black hole masses much greater than the Planck
mass, one important early work which captures the overall features of dynamical backreaction is
that by Bardeen \cite{Bardeen} and its further elaboration by  Masser \cite{Masser}.
 [See \cite{HRS} for more details.]

\subsection{Fluctuation-Dissipaton Relation for Black Holes}

Candelas and Sciama \cite{CanSci} were the first to suggest that the black hole radiance problem
can be understood as a quantum dissipative system. For a static black hole in equilibrium
with its Hawking radiation, Mottola \cite{Mot} used the formal equivalence to a thermal field
problem to show that in some generalized Hartle-Hawking  state a fluctuation-dissipation
relation (FDR) exists between the expectation values of  the commutator and anti-commutator of
the energy-momentum tensor of the scalar field, a form familiar in linear response theory
\cite{LRT}. In a recent essay Raval, Sinha and I \cite{HRS} showed how both of these proposals
are flawed. We showed why for a {\it bona fide} backreaction study of thermal radiance on a
quasi-static black hole, one should consider {\it ab initio} states more  general than the
Hartle-Hawking state.  To obtain a causal fluctuation-dissipation relation (FDR)  one needs to use
the in-in (or Schwinger-Keldysh) formalism applied to a class of quasistatic metrics
(generalization of York \cite{York}) and calculate the fluctuations  of the energy momentum
tensor for the noise kernel. So far we have \cite{CamHu} completed such a calculation only for
thermal fields in a weak gravitational field  which depicts the far-field limit of a Schwarzschild
black hole spacetime \cite{Bari}. For the noise kernel of
quantum fields near a Schwarzschild horizon Phillips \cite{Phillips}
has obtained a finite expression using the Gaussian approximation for the Green function.
 The accuracy of this approximation worsens near the horizon and a more
reliable calculation would require the inclusion of higher order terms in the
Schwinger-DeWitt expansion (the $a_1, a_2$ coefficients). In
the following we outline the recent result of Campos and Hu \cite{CamHu} 
for thermal fields in a weak gravitational background, which can be viewed
as the far-field limit of this problem.

\subsection{Thermal Fields in Black Hole Spacetimes}

The behavior of a relativistic quantum field at finite temperature in a weak gravitational field has
been studied before by a number of groups \cite{GPY,Reb,ABFT}
for scalar and abelian
gauge fields.  In these work, the thermal graviton polarization tensor and the effective action have
been calculated and applied to the study of the stability of hot flat and curved spaces 
and the ``dynamics" of cosmological perturbations. To describe screening effects and stability of thermal
(linearized) quantum gravity, one needs only the real part of the polarization tensor, but for
damping effects, the imaginary part is essential. The gravitational polarization tensor obtained
from the thermal graviton self-energy represents only a part (the thermal correction to the vacuum
polarization) of the finite temperature quantum stress tensor. There is in general also contributions
from particle creation (from vacuum fluctuations at zero and finite temperatures). These processes
engender dissipation in the dynamics of the gravitational field and their fluctuations appear as
noise in the thermal field. We have found such a relation between these two processes,
 which embodies the backreaction self-consistently.

Our calculation of the quantum corrections of the scalar field to the thermal graviton polarization
tensor was carried out by means of the Feynman-Vernon \cite{if} influence functional (IF).  It
yields results identical to that obtained before by means of linear response theory (LRT)
  \cite{Reb,ABFT}. From the IF one can obtain the noise and  dissipation  kernels explicitly which
satisfy a Fluctuation-Dissipation Relation (FDR) \cite{FDR} at all temperatures. 
This relation captures the essence of backreaction fully.

 We consider a free massless scalar field $\Phi$ arbitrarily coupled to a gravitational field
$g_{\mu\nu}$ with classical action (3).
In the weak field limit we consider a small
perturbation $h_{\mu\nu}$ from flat spacetime $\eta_{\mu\nu}$ in the form
$   g_{\mu\nu}(x) =  \eta_{\mu\nu} + h_{\mu\nu}(x)$
with signature $(-,+,\cdots ,+)$ for the Minkowski metric.
The CTP effective action at finite temperature $T= 1/ \beta$ for a free quantum scalar field in this
gravitational background is given by 
\begin{equation}
   \Gamma^\beta_{CTP}[h^\pm_{\mu\nu}]
        \ = \ S^{div}_g[h^+_{\mu\nu}] 
             -S^{div}_g[h^-_{\mu\nu}]
             -{i\over2}Tr\{ \ln\bar G^\beta_{ab}[h^\pm_{\mu\nu}]\},
   \label{eq:eff act two fields}       
\end{equation}
where $ a, b = \pm$ denote the forward and backward time path and $\bar
G^\beta_{ab}[h^\pm_{\mu\nu}]$ is the complete $2\times 2$ matrix propagator with 
thermal boundary conditions for the differential operator $\Box + V^{(1)} + V^{(2)} + \cdots$
where $V^{(n)}$ contain terms of $nth$ order in $h_{\mu\nu}$ from the expansion of the scalar
curvature in $S_f$.   Here $S_g^{div}$ is the (divergent) gravitational action
\begin{eqnarray}
   S^{div}_g[g_{\mu\nu}]
        & \ = \ & {1\over\ell^2_P}\int d^nx\ \sqrt{-g}R(x)
                \nonumber \\
        &       & +{\lambda \bar\mu^{n-4}\over4(n-4)}
                   \int d^nx\ \sqrt{-g}
                   \left[ 3R_{\mu\nu\rho\sigma}(x)
                           R^{\mu\nu\rho\sigma}(x)
                         -\left( 1-360(\xi - {1\over6})^2
                          \right)R(x)R(x)
                   \right].
\end{eqnarray}
The first term is the classical Einstein-Hilbert action and the second (divergent) term in four
dimensions is the counterterm introduced to renormalize the effective action. 
As before, $\ell^2_P = 16\pi G$, $\lambda = (2880\pi^2)^{-1}$ and $\bar\mu$ is an arbitrary mass scale.
(It is noteworthy that the counterterms are independent of the temperature because the thermal
contribution to the effective action does not contain additional divergencies.)

We skip the details \cite{CamHu} and quote the results. The noise and dissipation kernels
 are expressed in terms of the propagators $\tilde
G^\beta_{\pm\mp}$, (here tilde indicates the Fourier transform and the + - signs indicate the time
branches in CTP) respectively, as
\begin{equation}
   \kl{\tilde N}{}{k}
        \ = \  -{1\over4}\int {d^4q\over(2\pi)^4}\ 
                [ \tilde G^\beta_{-+}(k+q)\tilde G^\beta_{+-}(q)
                 +\tilde G^\beta_{+-}(k+q)\tilde G^\beta_{-+}(q)]
                \kl{T}{}{q,k},
   \label{eq:noise}
\end{equation}
\begin{equation}
    \kl{\tilde D}{}{k}
        \ = \  {i\over4}\int {d^4q\over(2\pi)^4}\ 
               [ \tilde G^\beta_{-+}(k+q)\tilde G^\beta_{+-}(q)
                -\tilde G^\beta_{+-}(k+q)\tilde G^\beta_{-+}(q)]
               \kl{T}{}{q,k},
   \label{eq:dissipation}
\end{equation}
It is easy to show that they are related by the thermal identity
\begin{equation}
   \kl{\tilde N}{}{k} 
        \ = \ i\coth\left({\beta k^o\over2}\right)\kl{\tilde D}{}{k}.
\end{equation}
In coordinate space we have the analogous expression
\begin{equation}
   \kl{N}{}{x} 
        \ = \ \int d^4x'\ \mbox{\rm K}_{FD}(x-x')\kl{D}{}{x'},
\end{equation}
where the fluctuation-dissipation kernel $\mbox{\rm K}_{FD}(x-x')$ is
given by the integral
\begin{equation}
   \mbox{\rm K}_{FD}(x-x')
        \ = \ i \int {d^4k\over(2\pi)^4}\
                     e^{ik\cdot(x-x')}
                     \coth\left({\beta k^o\over2}\right).
\end{equation}
Defining the variance of the energy momentum tensor of the thermal field
$ \hat t^{\mu\nu}_\beta (x) \equiv \hat T^{\mu\nu}(x) 
 -\langle \hat T^{\mu\nu}(x) \rangle_\beta \hat I$
one can show that 
\begin{equation}
   \langle \{ \hat t^{\mu\nu}_\beta (x),
                   \hat t^{\rho\sigma}_\beta (x') \}   \rangle_\beta 
        \ = \  8\ \kl{N}{}{x-x'},
\end{equation}
\begin{equation}
   \langle [ \hat t^{\mu\nu}_\beta (x),
                   \hat t^{\rho\sigma}_\beta (x')]     \rangle_\beta 
        \ = \  8i\ \kl{D}{}{x-x'}.
\end{equation}
From the CTP effective action one can also derive an Einstein-Langevin equation governing the
evolution of the gravitational field under the dynamical influence of the thermal field, with a
stochastic source term whose autocorrelation is given by the noise kernel.  This is not so easily
obtainable by the conventional methods such as LRT in thermal field theory.

\section{Semiclassical Gravity as Mesoscopic Physics}

In the above I have sketched some current activities in stochastic gravity. As we have seen the
main issue in the stochastic regime is that of noise and fluctuations. Let us now explore its
implications. In particular, what can we say about quantum gravity now that this new theory is
supposedly one further step closer to it than semiclassical gravity. In order to answer this question
we need to examine where stochastic is placed between semiclassical and quantum in so far as the
main physical issues are concerned. We also need to discuss some philosophical issues related to
how we view the structure and origin of spacetime and re-examine the meaning of quantizing
gravity. For these we need first to ponder on the relation between quantum and classical as well as
micro and macro physics.

On this issue I have proposed to view the low energy theory (classical GR) as the hydrodynamic-
collective state of the substructures of spacetime. Only these basic constituents
 (most likely fermions)-- and not
the collective variables -- obey quantum mechanical rules. (One can quantize these
variables but they describe excitations of the collective modes such as phonons, plasmons etc,
 not the underlying basic consistituents such as atoms or electrons.)  In this view, quantum gravity does not
refer to a quantization of metric or connections (which describe the collective modes), but to the more
basic, as yet unknown (strings?) consituents. To see the effects of this deeper structure with its
coherence properties from the stochastic regime we suggest to rely on topological signatures, 
effective field theory and
the correlation hierarchy and its dynamics. We shall put aside topological 
considerations and only address in the next three sections  three groups of issues --
stochasticity, collectivity and correlations -- following the themes:
``semiclassical gravity as mesoscopic physics'' \cite{meso}, ``general relativity as
geometro-hydrodynamics" \cite{Sak,grhydro} and ``quantum microdynamics via correlation hierarchy"
\cite{dch,cddn,StoBol}.  I will spend less space on stochasticity even though it is the central
theme of this new regime of interest, because it has been discussed extensively in recent articles,
e.g, \cite{CH94,HuSin,HM3,CV96,CH97}. Rather I will expand on the other two issues 
and indicate productive avenues for further investigation.

In an essay written in 1994 \cite{meso} I proposed to examine some important issues in
semiclassical gravity in the light of mesoscopic physics: Issues such as the transition from
quantum to classical spacetime via decoherence, cross-over behavior at the Planck scale,
tunneling and particle creation, growth of density contrast from vacuum fluctuations, or finite size
effect in curved spacetime phase transitions, share some basic concerns of mesoscopic physics for
condensed matter, atoms or nuclei, in the quantum / classical and the micro / macro interfaces, or
the discrete / continuum and the stochastic / deterministic transitions.  We pointed out that
underlying these issues are three main factors: quantum coherence, fluctuations and correlation.
We discussed how a deeper understanding of these aspects of fields and spacetimes can help us
address some basic problems, such as Planck scale metric fluctuations, cosmological phase
transition and structure formation, and the black hole entropy, end-state and information paradox.

Mesoscopic physics deals with problems where the characteristic interaction scales or sample
sizes are intermediate between the microscopic and the macroscopic. For the experts they refer to
aspecific set of problems in condensed matter and atomic / optical physics (see, e.g.,
\cite{mesobooks}). For the present discussion, I will adopt a more general definition, with `meso'
referring to the interface between macro and micro on the one hand and the interface between
classical and quantum on the other. \footnote{ Another meaning of mesoscopia can be defined
with respect to structures and interactions. Instead of dwelling on these individual processes in
their specific context, one can refer to the general category of problems which exist in between
two distinct levels of matter structure or interaction scales, such as between the molecular and
atomic scales, the QED lepton-hadron, the nucleon and particle (quark-gluon)
scales, the QCD and GUT(grand unification  theory) scale (with or without deserts in-between),
and of course, from GUT to QG (quantum gravity) scale, which is depicted by semiclassical
gravity. The distinct levels of interaction are not arbitrarily picked, they obey theories of a
`fundamental' (QED, QCD) or derived (atomic, nuclear interaction) nature -- even what we today
view as fundamental interactions may just be collective states of a deeper structure. The meso
scales between them have common traits. They usually fall in the range where the approximations taken  from
either level (e.g., low energy QCD versus perturbative hadron physics)
fail, and new structure depicted by new collective variables and new language are called for. 
The new problems encountered in condensed matter and nuclear/particle
physics fall under such a conceptual category, so do the problems of extending semiclassical
gravity towards quantum gravity or projecting quantum gravity (e.g., superstring theory) onto
low energy particle physics (the standard model).}
These two aspects will often bring in the continuum / discrete and the deterministic / stochastic
factors.  I showed  how issues concerning the micro / macro interface and the quantum to
 classical transition arise in quantum cosmology and semiclassical gravity in a way categorically
similar to the new problems arising from condensed matter and atomic / optical physics (and, at a
higher energy level, particle/nuclear physics, at the quark-gluon and nucleon interface). Many
issues are related to the coherence and correlation properties of quantum systems, and involve
 stochastic notions, such as noise, fluctuations,
dissipation and diffusion in the treatment of transport, scattering and propagation processes. The
advantage of making such a comparison between these two apparently disjoint disciplines is two-fold:
 The  theory of mesoscopic processes which can be tested in laboratories with the newly
developed nanotechnology can enrich our understanding of the basic issues common to these
disciplines while being extended to the realm of general relativity and quantum gravity. The formal
techniques developed and applied to problems in quantum field theory and spacetime geometry
can be adopted to treat condensed matter and atomic/optical systems with more rigor, accuracy
and completeness. Many conceptual and technical challenges are posed by mescoscopic processes
in both areas.

\subsection{Mesoscopic Physics -- Fundamental Issues at the Quantum / Classical and 
Micro /  Macro Interfaces}

Viewing in a more theoretical light, we can decipher three aspects which underlie all mesoscopic
processes, in gravitation and in condensed matter physics. They are quantum coherence,
fluctuations and correlations. They manifest in the quantum - classical and the micro - macro interfaces. \\

\noindent 1. Fluctuations and Decoherence\\

\noindent
Fluctuations and noise in the environment are responsible for decoherence in the system, which is
a necessary condition for quantum to classical transition \cite{conhis,envdec}. Classical
description in terms of definite trajectories in phase space requires correlations between conjugate
variables. Noise and fluctuations destroy this correlation. The observed classical reality as an 
emergent phenomenon from quantum description has intrinsic stochastic behavior
\cite{GelHar2,Tsukuba}.\\

\noindent 2. Coherence and Dissipation \\

This is the counterpart to the above, as fluctuations and dissipation are balanced by the
fluctuation-dissipation relation.  The degree of coherence here refers to the phase information in a
quantum system which can be corrupted by its interaction with an environment,
resulting in a stochastic classical dynamics for the system.  Coherence in quantum systems is
altered by  dissipative effects, as occurs in macroscopic quantum phenomena \cite{CalLeg83},
e.g., in tunneling with dissipation at finite temperature.\\

\noindent 3. Correlation and Collectivity\\

A useful signifier of the statistical properties of a system is its correlation functions -- the BBGKY
hierarchy in classical physics, or the Schwinger-Dyson equations in quantum field theory. It can be
used to measure the degree of coherence in either  the classical (correlation of the wave functions
in space and time) or the quantum senses (phase information). An example of collectivity is the
hydrodynamic variables versus the micro-variables: the transition from kinetic theory to
hydrodynamics is well-known. The formal treatment refers to deriving the Naviers-Stoke equation
from the BBKGY hierarchy. Thus lies the relation of correlation and collectivity which manifests in
the micro to macro transition. Combined with the consideration of noise and decoherence above
we can see that the quantum / classical and the micro / macro transitions are interrelated issues. \\

\subsection{Effective Theories: Renormalizability, Stochasticity and Collectivity}

The same factors arise in effective theories, which are theories valid at a lower energy or a larger
scale but constructed or derived from more fundamental theories for the more basic constituents.
An examples is the Fermi 4 point interaction as a low energy limit of the Weinberg-Salam
electroweak interaction. Important issues are 1) whether the low energy effective theory is
renormalizable, or effectively renormalizable -- deeper understanding of effective theories has
changed our view on renormalizability (see e.g., \cite{Weinberg}). 2) How do the effects of the
high energy sector or processes at a shorter lengthscale show up, if at all, at a  larger scale in the
low energy observation range? 3) Usually the low energy physics is described by a different set of
variables from the high energy physics --what we call the collective variables. How do we
construct the collective state from the microphysics? An even harder question: If we only know
about the dynamics of the collective state, how much information about the deeper structure can
we infer? 4) The interphase  between high and low energy sectors can involve a cross-over or a
phase transition -- what determines its character?  In particular, fluctuations carry important
information about the interphase and if it persists even at miniscule amount, can provide valuable
information about the short scale behavior.  In selected conditions such as in inflationary universe
(as a `zoom lense' \cite{cgea}) or black holes (as a `microscope' \cite{JabVol})  it offers hope
to probe into sub-Planckian physics through structure formation \cite{CH95} or Hawking
radiation processes \cite{JacUnr}.

The above issues were phrased in a way which are particularly relevant to the search for a viable
theory of quantum gravity from low energy physics -- by this we mean a quantum theory for the
substructure of spacetime, not the quantization of general relativity. On the issue of stochasticity,
Calzetta and I \cite{CH97} have studied an effective field theory and came up with a better
understanding of the threshold behavior. We explored how the existence of a field with a heavy
mass influences the low energy dynamics of a quantum field with a light mass by expounding the
stochastic characteristics of their interactions which take on the form of {\it fluctuations} in the
number of (heavy field) particles created at the threshold, and {\it dissipation} in the dynamics of
the light fields, arising from the backreaction of produced heavy particles. We claim that the
stochastic nature of effective field theories is intrinsic, in that dissipation and fluctuations are
present both above and below the threshold. Stochasticity builds up exponentially quickly as the
heavy threshold is approached from below, becoming dominant once the threshold is crossed. But
it also exists below the threshold and is in principle detectable, albeit strongly suppressed at low
energies. The results derived here can be used to give a quantitative definition of the
`effectiveness' of a theory in terms of the relative weight of the deterministic versus the stochastic
behavior at different energy scales.

In addition to stochasticity, one needs also to pay attention to two sets of issues 
  a) How {\it collective variables} can be assigned for low energy physics \cite{Spain}. For
gravity, if  we assume that metric or connection are the collective variables, how are they
derivable from a deeper structure (e.g., strings) without us knowing the details of their
interactions (e.g., string field theory)?.
b) Viewing general relativity as the  {\it hydrodynamic limit} of quantum gravity, examine the
equivalent of the kinetic theory regime \cite{grhydro}. Work on decoherent history of hydrodynamic
variables by Gell-Mann, Hartle and Halliwell \cite{dechyd}, and on correlation
history by Calzetta and Hu \cite{dch,cddn} will be useful for pursuing these ideas.  I will
expand on the collectivity aspects in the next two sections before turning to the correlation
aspect. \footnote{The next two sections are excerpted from (the unpublished part of) 
an article \cite{grhydro} 
Readers prone to be bored or annoyed by  philsophical discourses 
should proceed to Sec VI.}

\section{General Relativity as Geometro-Hydrodynamics}

In an essay written in 1996 \cite{grhydro} for the Second International Sakharov Conference, in the
spirit of his `metric elasticity' idea \cite{Sak}, I presented the viewpoint that general
relativity is hydrodynamics. It describes the  collective state (call it `spacetons'?) 
of a system of strongly interacting
quantum objects (strings?) which span the spacetime substructure . We
examined the various conditions which underlie the transition from some candidate theory of
quantum gravity to general relativity, specifically, the long wavelength, low energy (infrared)
limits, the quantum to classical transition, the discrete to continuum limit, and the emergence of a
macroscopic collective state from the microscopic consitituents and interactions of spacetime and
fields. In the `top-down' approach, we argued that nonequilibrium quantum field theory is needed
in showing how general relativity arises as various limits are taken in all candidate theories of
quantum gravity, such as string theory, quantum geometry (via the Ashtekar spin connections or
the Rovelli-Smolin loop representations), and simplicial quantum gravity.  In the `bottom-up'
approach, which is the path we have taken,  one starts with the semiclassical theory of gravity and
examines how it is modified by graviton and quantum field excitations near and above the Planck
scale. We mentioned three aspects based on our recent findings:
1) Emergence of stochastic behavior of spacetime and matter fields depicted by an
Einstein-Langevin equation. The backreaction of quantum fields on the classical background
spacetime manifests as a fluctuation-dissipation relation (discussed above).
2) Manifestation of stochastic behavior in effective theories below the
threshold arising from excitations above. The implication for general relativity is that such
Planckian effects, though exponentially suppressed, is in principle detectable at sub-Planckian
energies \cite{CH97}. 3) Decoherence of correlation histories
and quantum to classical transition \cite{dch}. Following the observation of  Gell-Mann and
Hartle that the hydrodynamic variables which obey conservation laws are most readily decohered,
we showed in \cite{grhydro}
 how one can, in the spirit of Wheeler \cite{WheelerBianchi}, view the conserved Bianchi identity
obeyed by the Einstein tensor as an indication that general relativity is a geometry-hydrodynamic theory.

\subsection{`Top-down': How to reach the correct limits}

The possible transitions we expect to find between quantum gravity and general relativity, i.e.,  quantum to
classical transition, low energy, long wavelength (infrared) limits, discrete to continuum limit,
extended structure to point structure, and micro/constituents versus  macro/collective states,
manifest in varying degrees of transparency in three leading types of candidate theories of
quantum gravity: the superstring theory \cite{string}, the loop representation of quantum
geometry via spin connections \cite{loop}, and simplicial quantum gravity \cite{simpQG}. In
string theory, a spin-two particle is contained in the string excitations, and it is easy to see
the limit taken from an extended structure to a point. The larger problem of how the target space
(e.g., spacetime of 26 dimensions for bosonic string) can be deduced from, or at least treated on
the same footing as, the world-volume of fundamental branes, still remains elusive. The
Bekenstein-Hawking expression for the black hole entropy \cite{BekHaw} originally derived in
semiclassical gravity is obtained as the tree level result of many quantum theories of gravity
\cite{tdbhent}. \footnote{Jacobson \cite{JacEqState} has used the thermodynamic expression for
black hole entropy to show how  Einstein's equation can be derived as a thermodynamic equation
of state. The underlying philosophy of this view is similar to ours.}  But in the construction of a
statistical mechanical entropy  \cite{smbhent} from quantum field theory in curved spacetime, it is
not so clear which of the many internal degrees of freedom of string excitations contribute to the
leading quantum correction term. It is encouraging that recent advances in D-brane technology
and duality relations have provided a statistical mechanical origin of black hole entropy from
string theory albeit so far only for near-extremal black holes \cite{strbhent}. 
This linkage with low energy physics (semiclassical gravity
results) will illuminate on how the collective variables are chosen and the collective state formed.
In the quantum relativity approach using Ashtekar's spin connection and Rovelli-Smolin's loop
representation, the picture of a one-dimensional quantum weave behaving like a polymer is
evoked \cite{AshErice}. When viewed at a larger scale the weaves appear to `knit' a higher
dimensional spacetime structure. This is an interesting picture, but how this collective process
comes about -- i.e., how the physical spacetime becomes  a dynamically preferred entity
and an infrared stable structure -- remains to be explicated (cf. protein-folding?). In simplicial
quantum gravity \cite{simpQG}, the classical limit might be obtained more easily in some versions
(e.g., in the Ponsano-Regge 6j calculus \cite{PonReg}, it is quite similar to the treatment of
ordinary spin systems via group-theoretical means, in place of the more involved considerations of
environment-induced decoherence \cite{envdec}), but essential properties like diffeomorphism
invariance in the continuum limit are not guaranteed, such as in Regge calculus.  Dynamical
triangulation procedure \cite{dyntriQG} was believed to work nicely in these respects. But there
are speculations  that a first order transition may arise which can destroy the long
wavelength niceties. How the general relativity limit comes about  is not yet fully understood.

Many structural aspects of these theories in their asymptotic regimes
(defined by the above-mentioned limits) near the Planck scale bear sufficient resemblance to the
physics in the atomic and nuclear scales
that I think it is useful to examine the underlying issues in the light of
these better-understood and well-tested theories. These include on the one
hand theories of `fundamental' interactions and constituents, such as
quantum electrodynamics (QED), quantum chromodynamics (QCD) --
add to them the well-developed yet untested theories of supersymmetry
(SUSY) and grand unified theories (GUT) -- which are indeed what piloted
many of today's candidate theories of quantum gravity, and on the other hand theories about how
these interactions
and constituents manifest in a collective setting -- theories traditionally discussed in condensed
matter physics using methods of
statistical mehanics and many-body theories.
These two aspects are not disjoint, but are interlinked in any realistic
description of nature (see \cite{HK,Spain}). They should be addressed
together in the search for a new theory describing matter and spacetime
at a deeper level.  The collective state description has not been emphasized as much as the
fundamental interaction description. We call  attention to its relevance because especially in this
stage of development of candidate theories of quantum gravity, 
deducing their behavior and testing their consequences at low energy  constitute an
important discriminant of their viability. Low energy particle
spectrum and black hole entropy are prime examples among the currently pursued topics.

Take, for example, the interesting observations related above, that four-dimensional spacetime  is
an apparent (as observed at low energy)
rather than a `real' (at Planck energy scale) entity --  highlighted in Susskind's
\cite{Sus} world as hologram  and 't Hooft's
\cite{tHo} view of the string theoretical basis of black hole dynamics and
thermodynamics. General relativity could be an emergent theory in some `macroscopic',
averaged sense at the low energy, long wavelength limit.
The fact that fundamental constituents manifest very different features at
lower energies is not so surprising, they are encountered in
almost all levels of structure -- molecules from atoms, nuclei from quarks --
referred to categorically as `collective states'.
How relevant and useful these variables or states are depend
critically on the scale and nature of
the physics one wants to probe. One cannot say that one is better than the
other without stipulating the range of energy in question, the nature of
the probe or the precision of the measurement. Just as thermodynamic variables are powerful and
economical in the description of long wavelength processes, they are
completely useless at molecular scales. Even in molecular kinetic theory,
different variables (distribution and correlation functions) are needed for different ranges of interactions. 
In treating the relation of quantum gravity to general relativity it is
useful to bear in mind these general features we learned from
more familiar processes.

Even when one is given the correct theory of the constituents, it is not
always an easy task to construct the appropriate collective variables
for the description of the relevant physics at a stipulated scale.
Not only are the derived structures different from their constituents, their effective interactions
can also be of a different nature.
There used to be a belief (myth) that once one has the fundamental theory it is only a matter of
details to work out an effective theory for its lower-energy counterparts. Notice how nontrivial it
is to deduce the nuclear force from quark-gluon interactions, despite our firm knowledge that
QCD is the progenitor theory of nucleons and nuclear forces.
Also, no one has been clever enough to have derived, say,
elasticity from QED yet. Even if it
is possible to introduce the approximations to derive it, we know it is plain
foolish to carry out such a calculation, because at sufficiently low energy,
one can comfortably use the stress and strain variables for the description of
elasticity. (Little wonder quantum mechanics, let alone QED, is not a required
course in mechanical engineering.)

\subsection{`Bottom-Up': Tell-tale signs from low energy}

How the low energy behavior of a theory is related to its high energy behavior
(issues of effective decoupling and renormalizability naturally would arise
\cite{eft}), whether one can decipher traces
of its high energy interactions or  remnants of its high energy components,
have been the central task of physics since the discovery of atoms in the last
century and subatomic particles in this century to today's attack on unified
theories at ultrahigh energy.
The symmetry of the particles and interactions existing at low energies
are the only raw data we can rely on to construct (and appraise the degree of success of)
a new unified theory. (Such is the central mission of e.g., 
string phenomenology in reproducing the low energy particle spectrum.)
Some salient features of general relativity such as diffeomorphism invariance,
Minkowsky spacetime as a stable ground state, etc., are necessary
conditions for any quantum theory of gravity to meet at the low energy limit.
\footnote{Note that if we view general relativity as a hydrodynamic theory 
in the same sense as the nuclear
rotational and vibration states in the collective or liquid drop model, we can see that as much as
the symmetries of rotational and vibrational motion provide a useful description of the large scale
motion of a nucleus, they have no place in the fundamental symmetries of nucleons,
much less their constituents, the quarks and gluons. In this sense one could
also question the neccesity and legitimacy of basic laws like Lorentz
invariance and diffeomorphism invariance at a  more fundamental level. It
should not surprise us if they no longer hold for trans-Planckian physics.}
Approaching Planck energy from below,  the beautifully simple yet deep theory of
black hole thermodynamics \cite{BekHaw} first discovered in semiclassical gravity
is serving as a guide and providing a checkpoint for viable quantum gravity theories. 
Concerning the nature of the legacy (actually, the `leftovers' ) 
from the physics at high energy, or special tell-tale signs  at low energy, I would suggest paying careful
attention to two features:  topology and stochasticity. Topology refers to both nontrivial
spactimes and field configurations while stochasticity refers to the
 coarse-grained remnants of microphysics  and fluctuation effects at the cross-over.
 Here we will only focus on the latter feature, which is the central theme of stochastic gravity.

\subsubsection{Fluctuations and noise at the threshold}

An important feature of physics at the Planck scale depicted by semiclassical
gravity is the backreaction of  quantum effects
of particles and fields, such as vacuum polarization and particle creation,
on the classical gravitational spacetime. This is an essential step beyond
classical relativity for the linkage with quantum gravity.
For example, generalization to the $R + R^2$ theory of gravity
is a necessary product from the renormalization considerations of
quantum field theory in curved spacetimes. It should also be the low energy
form of string theory (plus dilaton and antisymmetric fields).
Backreaction demands more, in that the quantum matter field is solved
consistently with the classical gravitational field \cite{cpcbkr}.
The consistency requirement in a backreaction calculation brings in two
new aspects:\\
1) The classical gravitational field obeys a dynamics which contains a dissipation
component arising from the backreaction of particle creation in the quantum
field. The dissipation effect is in general nonlocal, as it is influenced by
particle creation not only occuring at one moment, but also integrated over the
entire history of this process \cite{HarHu,CH87}.\\
2) Creation of particles in the quantum matter field at the Planck energy
(which is responsible for the dissipative dynamics of the gravitational field)
can be depicted as a source which has both a deterministic and a stochastic
component. The first part is the averaged energy density of created particles,
which is known in previous treatments. The second part measures
the difference of the amount of particles created in two neighboring histories
and is depicted by a nonlocal kernel, the correlator of colored
noise \cite{CH94,HM2}.  The dissipation and noise kernels are related by a
fluctuation-disspation relation.  As described above, the backreaction equation 
is in the form of a Langevin equation, which we call
the Einstein-Langevin equation \cite{HM3,HuSin}.

The Einstein-Langevin equation constitutes a new frontier for us to explore
possible phase transition and vacuum instability issues, which we believe many
of the `top-down' approaches would also encounter in this cross-over regime.

\subsubsection{Stochastic behavior below the threshold}

What are the tell-tale signs for a low energy observer of the existence of
a high energy sector in the context of an effective field theory?
We wish to adopt an open system viewpoint to consider effective theories and explore
their statistical mechanical properties.
The question is to compare the difference between a
theory operative, (i.e., giving an adequate description) at low energies
(as an open system, with the high energy sector acting as the environment)
to an exact low energy theory taken as a closed system.
We know that there are subtle differences between the two, arising from the 
backreaction of the heavy  on the light sector.
Though not obvious, the stochastic behavior associated with particle creation
above the threshold (which for gravitational processes is the Planck energy)
is related to the dissipative behavior of the background spacetime dynamics.
(This was known for some time, see, e.g.,  \cite{Physica}.) 
Schwinger's result \cite{Sch}
for pair production in a strong electromagnetic field 
is a well-known example.  This effect at very low energy has however been ignored,
as it is usually regarded as background noise covered by very
soft photons. That such a noise carries information about the field at high
energy was only pointed out recently  \cite{CH97}.
Using a simple interacting field model, Calzetta and I found 
that even at energy way below the threshold, stochastic effects, albeit at
extremely small amplitudes, can reveal some general (certainly not the 
specific) properties of the high energy sector. Finally one can also show from the decoherence
aspects of quantum theories in reaching their classical limits \cite{grhydro} why general relativity can
be viewed as the hydrodynamic limit of quantum gravity.

\section{Low energy collective state physics and beyond}

Suppose one takes this viewpoint seriously, what are the possible implications? 
We can make a few general observations here.

\subsection{Quantizing metric may yield only phonon physics}

First, the laws governing these collective variables are classical,
macroscopic laws. It may not make full sense
to assume that by quantizing these variables directly
one would get the micro-quantum basis of the macro-classical theory,
as has been the dominant view in quantum gravity.
Just as the energy density $\rho$ and momentum densities $p$ in
the Einstein equation 
are the hydrodynamic variables of a matter field,
quantization should only be performed on
the microscopic fields $\Phi (x)$  from which they are constructed.
If one did so for the metric or the connection variables, one would
get the quantum excitations of geometry in the nature of phonons in relation to
atoms (or other quantum collective excitations in condensed matter physics).
That may be the next order of probe for us, and may reveal some interesting phenomena,
but it is still very remote from seeing the nucleon structure
in the solid lattice or the attributes of quantum electrodynamics.
In the analogy we mentioned above, we don't expect quantum elasticity to
tell us much about QED.

Second -- and this is perhaps the more interesting aspect -- assuming that
the metric and connections are the collective variables, from the way they
are constructed,  what can one say about their microscopic, quantum basis?
Historically this question was asked repeatedly when one probes from low to
high energy scales, trying to decipher the microscopic constituents and laws of
interactions from macroscopic phenomena. This is like going from phonons to
the structure of atoms, from nuclear rotational spectrum to nucleon strucuture
-- not an easy question to answer. But there are nevertheless ways to guide us,
e.g., in terms of some tell-tale signs.
In the above analogies, recall that atomic spectroscopy reveals many
properties about the electron-electron and electron-nucleus interactions,
low temperature anomalous behavior of specific heat reveals the quantum
properties of electrons, the intermediate boson model bridges the symmetry
of the collective modes with that of the independent nucleons.
To address questions like these, one needs to proceed from both ends:
One needs to postulate a theory of the microscopic structure, and work out
its collective states at large scale and low energies.
One also needs to comb through the consequences of the known low energy theory,
paying attention to subtle inconsistencies or mistakenly ignored trace effects
from higher energy processes. Indeed, this is what is going on today,
with string theory as the micro theory, and semiclassical gravity and 
particle phenomenology  as its low energy limit.
The viewpoint we are proposing would suggest focusing on collective
states (solitons?) of  excitations of the fundamental string on the one hand
and a detailed study of the possible new phenomena in quantum field theory
in curved spacetime on the other, such as flucutations and phase transitions around the
Planck energy, quantum corrections to the black hole entropy, resonance states and spacetime foams.

\subsection{Common features of collective states built from different constituents}

As mentioned above, there are two almost orthorgonal perspectives in
depicting the structure and properties of matter. One is by way of its
constituents and interactions, the other according to its collective behavior.
The former is the well-known and well-trodden path of discovery of QED, QCD,
etc. If we regard this chain of QED - QCD - GUT - QG as a vertical
progression depicting the hierarchy of basic constituents, there is also a
horizontal progression in terms of the stochastic - statistical
- kinetic - thermodynamic/hydrodynamic depiction of the collective states.
It should not surprise us that there exist similarites between matters in
the same collective state (e.g., hydrodynamics) but made from different
constituents.
Macroscopic behavior of electron plasmas are similar in many respects to the
quark-gluon plasma. Indeed, one talks about magneto-hydrodynamics from Maxwell's
theory as well as magneto-chromo hydrodynamics from QCD. In this long wavelength,
collision-dominated regime, they can both be depicted by the hydrodynamics of
fluid elements, which are governed simply by Newtonian mechanics.
The underlying
micro-theories are different, but the hydrodynamic states of these constituents
are similar. Here we are proposing that general relativity being the hydrodynamics limit
 (of some candidate theory of quantum gravity) is an effective theory
in the way that nuclear physics is with regard to QCD, and atomic physics is
with regard to QED. They are all low energy collective states
of a more fundamental set of laws and can share some similarities. The macroscopic,
hydrodynamic equations
and their conservation laws like the Naviers-Stoke and the continuity
equations are all based on dynamical and conservation laws of
microphysics (e.g., Newtonian mechanics), but when expressed in terms of the
appropriate collective variables, they can take on particularly simple and
telling forms. Thermodynamic variables like temperature, entropy, etc.
(think black hole analogy -- mass, surface area) are derived quantities
with their specific laws (three laws)
traceable via the rules of statistical mechanics (of Gibbs and Boltzmann)
to the laws of quantum mechanics.  Rules of statistical mechancis are important
when we probe into a deeper layer of structure from known low energy theories
such as semiclassical gravity: we need to know how to disentangle the collective
states in order to see how the microphysics works. \footnote{Savour the importance
of, say, coming up with a statistical mechanical definition of temperature
in a canonical ensemble as the rate of change of the accessible states of a
system in contact with a heat reservoir with respect to changes in energy, and
we can appreciate the importance of Gibbs' work in relation to quantum physics.}
It is hard to imagine how a complete theory of microphysics can be attained
without going through this step.

\subsection{Hydrodynamic fluctuations applied to black holes and cosmology}

A problem where this analogy with collective models may prove useful
is that of black hole entropy.
If we view the classical expression for black hole entropy to be a
hydrodynamic limit, and the corrections to it as arising from hydrodynamic
fluctuations, one could use linear response theory to approach conditions
near thermodynamic equilibrium and construct a non-equilibrium theory of
black hole thermodynamics.\footnote
{Black hole backreaction problem has been studied by many authors
before, notably by York \cite{York}, Anderson and Hiscock \cite{AndHis} and
their collaborators. We are taking a non-equilibrium statistical field theory
approach. 
We aim to get the fluctuations of the energy momentum tensor of a quantum
field in a perturbed Schwarzschild spacetime \cite{HPR},
examine how they might induce dissipations of the event horizon and 
deduce a susceptibility function of the black hole.
This would realize the proposal of Sciama that a black hole
in equilibrium with its Hawking radiation can be depicted as a 
quantum dissipative system \cite{CanSci}. (See also \cite{Mot})}
It also seems to us that many current attempts to
deduce the quantum corrections of black hole entropy
from the micro-quantum theory of strings
could be missing one step. This is like the correpondance between results predicted
from the independent particle (nucleon) model (where one can construct
the shell structures), and that from the liquid drop model (where one can
construct the collective motions) -- a gap exists which cannot easily be
filled by simple extensions of either models operative in their
respective domains of validity. This involves going from  the individual
nucleon wavefunctions to the collective states of a nucleus.
It is likely that only specific
appropriate combinations of fundamental string excitation modes
which survive in the long wavelength limit can contribute to the excitations
of the collective variables (area and surface gravity of black hole)
which enter in the (semiclassical gravity) black hole entropy.
\footnote{This statement made in July 1996 should be viewed in the context
of new developments since then in the statistical mechanical origin 
of black hole entropy via D-branes \cite{strbhent}}

Viewing classical GR as hydrodynamics stochastic gravity and
 Einstein-Langevin equation would depict the hydrodynamic fluctuations of spacetime dynamics
as induced by quantum field processes at the Planck scale .
One could study the behavior of metric and field
fluctuations with this Langevin equation in a way similar to that of
critical dynamics for fluids and condensed matter.

In summary, we note that progress of physics --
the probing of the structure and dynamics of matter
and spacetime -- has always moved in the direction from  low to high energies.
One needs to pay attention to the seemingly obvious facts at low energies
and probe into any discrepancy or subtlties not usually observed to find
hints to the deeper structures.
By examining how certain common characteristics of all successful low energy
theories  (here, we only discuss the hydrodynamic and thermodynamic aspects)
may recur in a new theory at a higher energy, and how they differ, we can
perhaps learn to ask the right questions and focus on some hitherto neglected aspects.\\

\section{Quantum Microdynamics via Correlation Hierarchy}

In the last two sections we touched on two issues deemed important in the
  transition period between quantum and classical gravity, i.e., 
stochasticity (or flucutuations) at the intermediate regime
 (stochastic semiclassical gravity) and collectivity at the low energy (general relativity) regime.
We now focus on the correlation aspect, which we think is important for probing the full quantum
regime. Along the way we will mention a few problems which may 
shed light on the passage from stochastic to quantum gravity.

\subsection{Correlation and Coherence}

If we look back at the equations in Sec. ID and compare the semiclassical (sC), stochastic (St) and 
quantum (Q) regimes we see first that in the sC case, the classical metric correlations is given by
the product of the vacuum expectation value of the energy momentum tensor whereas in the
quantum case the quantum average of the correlation of metric (operators) is given by the
quantum average with respect to the fluctuations in both the matter and the gravitational fields. In
the stochastic case the form is
 closer to the quantum case except that now the quantum average is replaced by the noise
average, and the average of the energy momentum tensor is taken with respect only to the matter
field. The important improvement over the semiclassical case is that it now carries information on the
correlation of the energy momentum tensor of the fields and its induced metric fluctuations. This
is another way to see why the stochastic description  is closer to the quantum truth. More
intuitively, the difference between quantum and semiclassical is that the latter loses all the coherence in
the quantum gravity sector. Stochastic improves on the semiclassical situation in that partial
information related to the coherence in the gravity sector is preserved as is reflected in the
backreaction from the quantum fields and manifests as induced metric fluctuations. That is why
we need to treat the noise terms with maximal respect. It contains quantum information absent in
the classical. The coherence in the geometry is related to the coherence in the matter field, as the
complete quantum description should be given by a coherent wave function of the combined
matter and gravity sectors. Since the degree of coherence can be measured in terms of
correlations our strategy is to examine the higher correlations of the matter field, starting with the
variance of the energy momentum tensor in order to probe into or retrieve whatever partial
coherence remains in the quantum gravity sector. The noise we worked out in the Einstein-Langevin
equation above contains the 4th order correlation of the quantum field (or gravitons
when considered as matter source) and  manifests as induced
metric fluctuations. Let us see what can be done to get closer to the quantum picture. 

If we view classical gravity as an effective theory, i.e., the metric or connection functions as
collective variables of some fundamental particles which make up spacetime  in the large and
general relativity as the hydrodynamic limit, we can also ask if there is a mid-way weighing station
like kinetic theory from molecular dynamics, from quantum micro-dynamics to classical
hydrodynamics. This transition involves both the micro to macro transition and the quantum to
classical transition, which is what constitutes the mesoscopic regime for us.

For our present purpose we can represent quantum gravity  as an interacting quantum field (of
fermions?) and we shall traverse this passage using the correlation dynamics from the (nPI) master
effective action. There are two aspects in this problem:  coherence of a field as measured by its
correlation (for quantum as well as classical), and quantum to classical transition. We wish to
treat both aspects with a quantum version of the correlation (BBGKY) hierarchy, the 
Schwinger-Dyson equations. There are three steps involved: First,
show how to derive the kinetic equations from quantum field theory -- or to go from Dyson to
Boltzmann \cite{CH88}. Second,
show how to introduce the open system concept to the hierarchy. For this we need to introduce
the notion of `slaving' in the hierarchy, which renders a subset made up of a definite number of
lower order correlation functions as an effectively open system,
 where it interacts with the environment made up of the higher correlation functions. Third, show
why there should be a stochastic term in the Boltzmann equation when contributions from the
higher correlation functions are included. 

\subsection{Kinetic field theory via master effective action}

The first step was taken in the 80's,  when  Calzetta and I  \cite{CH88}, 
amongst others \cite{kft80} (see \cite{qustme} for earlier work and \cite{kft90} for recent developments)
 showed how the quantum 
Boltzmann equation arises as a description of the dynamics of quasiparticles in the kinetic limit of
quantum field theory. The main element in the description of a nonequilibrium quantum field is its
Green functions, whose dynamics is given by the Dyson equations.
 For the second step, we showed in 1995 \cite{cddn}  how the Schwinger-Dyson equations can be obtained
from an $n=(\infty) PI$ master effective action and how the coarse-grained (truncation with
slaving) n-point correlation functions behave like an effectively open system. Recently
\cite{StoBol} we have taken the third step in  identifying such a noise term in the Boltzmann
equation (its classical limit reproduces the result of Kac and Logan \cite{KacLog}), and proving a
fluctuation-dissipation relation for these correlation noises. The resultant stochastic  Boltzmann
equation has features of both the Langevin and Boltzmann equations. 
With this one can then begin to investigate the possibility of using the
 correlation hierarchy to infer the quantum microdynamics.
 For illustration, we will just show the lowest order in the correlation hierarchy
by way of the master effective action.

The mean field and the two point function which one uses to deduce kinetic theory or critical
dynamics results are just the lowest two elements in the full Schwinger-Dyson (SD) hierarchy of
correlation functions.  In general the complete set is required to recover full (including phase)
information in a quantum field. If we now view the problem in this light we can see how 
dissipation and fluctuations arise when the hierarchy is truncated and the higher correlations are
slaved (we refer to these two procedures as coarse-graining), in the same way as how Boltzmann
equation is derived from the BBGKY hierarchy. What is new in our current understanding is that
there should also be a noise term in addition to the collision term in 
the Boltzmann equation.

In \cite{cddn}
we have shown how this hierarchy of SD equations can be derived from  the
master ($\infty PI$)  effective action so here we will just show the form of the 2PI.
For a scalar field $\Phi(x)$ with classical action $S[\Phi]$ under an external source $J(x)$  the
generating functional $W\left[ J\right] $\cite{effaction} is given by

\begin{equation}
\exp \left\{ iW\left[ J\right] \right\} =\int D\Phi \;\exp \left\{ iS\left[
\Phi \right] +i\int d^4x\;J\left( x\right) \Phi \left( x\right) \right\},
\label{genfun}
\end{equation}
from which one can obtain the expectation value or mean field

\begin{equation}
\phi \left( x\right) =\left. \frac{\delta W}{\delta J}\right| _{J=0}.
\label{orpar}
\end{equation}
The effective action is the Legendre transform of $W$

\begin{equation}
\Gamma \left[ \phi \right] =W\left[ J\right] -\int d^4x\;J\left( x\right)
\phi \left( x\right), 
\label{efact}
\end{equation}
from which we obtain the equation of motion
\begin{equation}
\frac{\delta \Gamma }{\delta \phi }=0  \label{eqmot}.
\end{equation}

In a causal theory, we must adopt Schwinger's CTP formalism. The point $x$
may therefore lie on either branch of the closed time path ($a,b = \pm$), or equivalently
we may have two background fields $\phi ^a\left( x\right) =\phi \left(
x^a\right) $. The classical action is defined as

\begin{equation}
S\left[ \Phi ^a\right] =S\left[ \Phi ^1\right] -S\left[ \Phi ^2\right] ^{*},
\label{ctpclac}
\end{equation}
which automatically accounts for all sign reversals. We also have two sources
\[
\int d^4x\;J_a\left( x\right) \Phi ^a\left( x\right) =\int d^4x\;\left[
J^1\left( x\right) \Phi ^1\left( x\right) -J^2\left( x\right) \Phi ^2\left(
x\right) \right] ,
\]
and obtain two equations of motion

\begin{equation}
\frac{\delta \Gamma }{\delta \phi ^a}=0 .
 \label{ctpeqmot}
\end{equation}
These equations always admit a solution where $\phi ^1=\phi ^2=\phi
$ is the physical mean field. After this identification, they
become a real and causal equation of motion for $\phi $.

The functional methods we have used so far to derive the dynamics of the
mean field may be adapted to investigate more general operators. In
order to find the equations of motion for two-point functions, for example,
we add a nonlocal source $K_{ab}(x,x^{\prime })$\cite{pi2,CH88}

\begin{equation}
\exp \left\{ iW\left[ J_a,K_{ab}\right] \right\} =\int D\Phi ^a\;\exp
i\left\{ S\left[ \Phi ^a\right] +\int d^4x\;J_a\Phi ^a+\frac 12\int
d^4xd^4x^{\prime }\;K_{ab}\Phi ^a\Phi ^b\right\} .
 \label{tpigf}
\end{equation}
It follows that

\[
\frac{\delta W}{\delta K_{ab}\left( x,x^{\prime }\right) }=\frac 12\left[
\phi ^a\left( x\right) \phi ^b\left( x^{\prime }\right) +G^{ab}\left(
x,x^{\prime }\right) \right] .
\]
Therefore the Legendre transform, the so-called 2PI effective action,

\begin{equation}
\Gamma \left[ \phi ^a,G^{ab}\right] =W\left[ J_a,K_{ab}\right] -\int
d^4x\;J_a\phi ^a-\frac 12\int d^4xd^4x^{\prime }\;K_{ab}\left[ \phi ^a\phi
^b+G^{ab}\right]  \label{gamma2pi}
\end{equation}
generates the equations of motion

\begin{equation}
\frac{\delta \Gamma }{\delta \phi ^a}=-J_a-K_{ab}\phi ^b;\;\frac{\delta
\Gamma }{\delta G^{ab}}=-\frac 12K_{ab}  \label{funceqmot}
\end{equation}

This is an (n=2) example of the nPI effective action. When $n \rightarrow \infty$, this is known as
the master effective action (MEA).
The master effective action is a functional of the whole string of Green functions of a field theory
whose variation generates the Schwinger -Dyson hierarchy.
In \cite{cddn}  we 1) gave a formal construction of the master effective
action, 2) showed how truncation in nPI is related to loop expansion and 
3) how  `slaving' leads to dissipation. 

\subsection{Theoretical considerations: nPI, 1/N and loop expansions}

In the above we have defined the master effective action and showed its relation to the
Schwinger-Dyson hierarchy. From this one can establish a kinetic theory of nonlinear 
quantum fields, to derive the kinetic equations \cite{CH88} and to derive 
a correlation noise arising from the slaving of the higher correlation functions. The stochastic
Boltzmann equation \cite{StoBol} contains features which would enable us to make connection
with the stochastic equation in semiclassical gravity. This comes about from the following
consideration: the Boltzmann equation describes the evolution of one particle distribution function
driven by a 2 particle collision integral, and the stochastic Boltzmann equation incorporates the
contribution of a higher order correlation function. The Langevin equation was derived in the
framework of an open system, the noise arsing from coarse-graining the environment. Truncation
and slaving as carried out in the hierarchy yields an effectively open system and the master
effective action leads to the stochastic Boltzmann equation similar to the Langevin equation in an 
open system. (From here we can see at work the two major paradigms in nonequilibrium
statistical mechanics: the Boltzmann-BBGKY and the Langevin-Fokker-Planck descriptions.) 
The corresponding situation for interacting quantum fields can be applied to quantum gravity --
assuming that it can be represented  by some interacting quantum field -- and illuminate on how
one should proceed from the standpoint of stochastic gravity. We can get a handle on the
correlation of the underlying field by examining the hierarchy of equations, of which the
Einstein-Langevin describes only the lowest order correlations: The relation of the mean field to
the two point function, and the two point function to the four (variance in the energy momentum
tensor). One can in principle move higher in this hierarchy to decipher the higher correlation
contributions. Notice that we have only dealt with the correlation aspect, the quantum to classical
aspect remains. This can be treated by the decoherence of correlation histories 
discussed earlier  in \cite{dch}. 

While we are discussing formal matters, I should mention that it is worthwhile to also include the
large N expansion for comparison.  There exists a relation between correlation order and the loop
order \cite{cddn}. One can also relate it to the order in large N expansion. It has been shown that
the leading order 1/N expansion for an N-component quantum field yields the equivalent 
of semiclassical gravity \cite{HarHor}. The leading order 1/N approximation yields mean field
dynamics of the Vlasov type \cite{KME} which shows Landau damping which is intrisically 
different from the
Boltzmann dissipation.  In contrast the equation obtained from the nPI (with slaving)
contains 
dissipation and fluctuations  manifestly.  It is apparent that the next to leading order incorporates
interactions corresponding to coherent scattering of particles.
It would be of interest to think about the relation between semiclassical and quantum in 
the light of the higher 1/N expansions,
which is quite different from the scenario associated with the correlation hierarchy.

\subsection{Physical considerations: strongly correlated systems}

At this point it is perhaps useful to bring back the opening theme of our discussion, i.e.,
semiclassical gravity as mesoscopic physics and examine similar concerns.

To practitioners in condensed matter and atomic/optical physics, mesoscopia refers
to rather specific problems where, for example, the sample size is comparable
to the probing scale (nanometers), or the interaction time is comparable to the
time of measurement (femtosecond), or that the electron wavefunction correlated
over the sample alters its transport properties, or that the fluctuation pattern
is reproducible and sample specific. Take quantum transport.
Traditional transport theory applied to macroscopic structures are based on kinetic theory while
that for mesoscopic structures is usually based on
near-equilibrium or linear response approximations (e.g., Landauer-B\"utiker formula).
New nanodevice operations involve nonlinear, fast-response and far-from-equilibrium 
processes which are sensitive to the phases of the electronic wavefunction over the
sample size. These necessitate a new microscopic theory of  quantum transport. One serious
approach is using the Keldysh method in conjunction with 
Wigner functions (e.g., \cite{mesotrans}). It is closely related
to the closed-time-path formalism we developed for nonequilibrium quantum fields aimed for
similar problems in the early universe and black holes \cite{CH88}.

Now focusing back on the issue of correlations and quantum coherence while using the analogy
with mesosystems we see that what appears on the right hand side of the Einstein-Langevin
equation -- the stress-energy two point function is analogous to conductance which is given by the
current-current two point function. What this means is that we are really calculating the transport
function of (the matter particles as depicted by) the quantum fields. Following  Einstein's keen
observation that spacetime dynamics is determined by (while also dictates) the matter (energy
density), we expect that the transport function represented by the current correlation in the matter
(fluctuations of the energy density)  would also have a geometric counterpart and equal
significance at a slightly higher energy scale. The hydrodynamics analogy we gave earlier 
also makes sense here: Conductivity, viscosity and other transport functions are hydrodynamic 
quantities. For many practical purposes we don't need to know about
the details of the fundamental constituents or their interactions
to establish an adequate depiction of the low or medium energy physics, but can model them with
semi-phenomenological concepts (like mean free path and collisional cross sections). In the
mesoscopic domain the simplest kinetic  model of transport using these concepts are no longer
accurate. One needs to work  with system-environment models and keep the phase information of
the collective electron wave functions. When the interaction among the constituents gets stronger,
effects associated with the higher 
correlation functions of the system begin to show up.  Studies in strongly correlated systems
 are revealing in these regards \cite{mesobooks,mesotrans}. For example, 
fluctuations in the conductance -- from the 4 point function of the current -- carry important
information such as the sample specific signature and universality. Although we are not quite in a
position, technically speaking, to calculate the energy
 momentum 4 point function, thinking about the problem in this way may open up many
interesting conceptual possibilities, e.g., what does universal conductance fluctuations mean for
spacetime and its underlying constituents? In the same vein, I think studies of nonperturbative
solutions of gravitational wave scattering \cite{gravscatt} will also  reveal interesting
information about the underlying structure of spacetime (beyond the hydrodynamic realm). Thus,
viewed in the light of mesoscopic physics, with stochastic gravity we are really begining to probe
into the higher correlations of quantum matter and with them the associated excitations of the
collective modes in geometro-hydrodynamics.
\footnote{Of course, walking down  this pathway, it will still take a while before one sees  the
microscopic quantum picture -- the constituents of spacetime, like electrons in quantum transport.
One may indeed never see it, because one needs to seek a different 
set of variables for the  basic constituents different from those for the collecive modes. But as far
as what low-energy observers can decipher, these collective phenomena are all that one can
observe and the hydrodynamic quantities such as the transport functions and their derived 
constructs actually offer a  better set of
variables for their description since the equations and the physics are simpler.}

\section{Towards Quantum Gravity}

We now integrate what we have discussed in the above and enumerate possible activities at the
Planck scale, related to the three aspects of fluctuations, correlation and collectivity.

\subsection{Quantum Tunneling, Particle Creation and Phase Transition at the Planck Scale}

The Langevin equation description of semiclassical gravity opens up a new
horizon at the juncture of general relativity and quantum gravity theories in that it enables one to
examine the properties of fluctuations in the
quantum matter fields and their effect on the stability of the classical
spacetime structure.
 
At the Planck  scale when  quantum  effects  of  gravity  become
significant,  physical  laws as well as the structure of spacetime and matter may undergo
fundamental changes in form and content.
Many such changes could be the outcome of phase transitions. The study of
Planck scale phase transitions is thus of fundamental theoretical  value.
Near the Planck time  when  the  gravitational  field  is  strong and when
spacetime geometry changes drastically, vacuum particle production is
abundant, and any phase  transition  would likely be accompanied by particle production.
In treating Planck scale phase transitions,  not  only  is  the
effective potential ill-defined, because the  background  field  changes  in time, but the
background field splitting often assumed in the derivation  of the effective Lagrangian would
become ineffective (because the  background  field can change as much
as, and as fast as, the  fluctuation  fields). In such cases (or in cases where global properties of
spacetime like  boundary  or topology are involved), one would need to use non-perturbative
methods such as instanton solutions (in Euclidean formalism). 
However, to incorporate statistical processes one needs a
real time description. It is difficult to join these two worlds but if we can (say, using a  new  route 
via Langiven or Fokker-Planck or master equations) we will be able to deal with a wider range of
issues.  Phase transition in the form of spinodal decomposition as applied to defect formation is
currently under investigation \cite{Greg}. Here I would like to comment on the process of 
nucleation via quantum tunneling in relation to stochastic gravity.

\subsubsection{Tunneling and Particle Creation as Vacuum Decay}

Our view is that both tunneling and particle creation are manifestations of
 vacuum instability but with different set-up of boundary conditions for
  these two processes.
(The tunneling probability, or the probability for finding a pair of
particles created, are both given by the  imaginary part of the
effective action.)
While in  a  tunneling  problem  the  system transits from one  definite
(metastable)  state  to  another,  in  particle production (from dynamic
spacetimes) it is a continuous change from an initial vacuum  to
a final  vacuum, with inequivalent Fock spaces at all intermediate states.
One can formulate this problem first in the setting
of quantum mechanical potential scattering using Bogolubov transformations,
 and then in the effective action formalism via vacuum persistence amplitudes.
The advantage of this unifying view is that many aspects of tunneling can be
addressed by established methods of treating particle creation.

\subsubsection{Tunneling with Particle Creation:  Dynamics and Dissipation}

If particle production occuring during tunneling is not  strong  enough
to disrupt  the  tunneling  process,  one  can
treat this as a test-field problem. Rubakov \cite{Rub}
 first attempted this problem with a nonunitary
Bogolubov transformation  (this method we do not find so agreeable, see also
criticism by Vachaspati and Vilenkin \cite{VacVil}) .
We prefer to use a real time approach and treat particle creation in the fluctuation fields as 
parametric amplification by the background field (as one encounters in the post-inflation
reheating problem \cite{RH1}).
If the particle creation is so strong as to alter the tunneling process,
one needs to take the backreaction into consideration and solve 
the `dynamics' of 
tunneling and particle creation self-consistently.
Since particle creation can be viewed as a  dissipative  process,
this becomes a problem  of tunneling with dissipation \cite{CalLeg}.
One can apply stochastic  field  theory for its treatment where dissipation
and noise are manifest. Insofar as particle creation is a form of amplified
quantum noise, the interesting processes of  stochastic resonance and
noise-induced transitions could also shed light on this issue.


\subsubsection{Tunneling and Decoherence in Quantum Cosmology}

Vilenkin has proposed a tunneling boundary condition in quantum cosmology
in the so called `birth of the universe' scenario \cite{Vil}.
What is the effect of particle creation on the tunneling
wavefunction?  Would dissipation terminate the tunneling process and
give `still birth' of the universe? Does it make sense to talk about matter
`before' (Euclidean time!) the universe? One can investigate this
issue in the context of minisuperspace quantum cosmology
by studying the effect of
dynamics on quantum fluctuations (of matter fields and spacetimes) during
tunneling. A related problem is decoherence and tunneling:
Could vacuum fluctuations induce a quantum to classical transition
in the tunneling wave function of the universe, giving rise to
a semiclassical regime with desirable attributes which could generate
our own universe, or will dissipation alter the picture irrevocably?
One can  incorporate
results on dissipative tunneling into earlier studies of decoherence with
backreaction in quantum cosmology (e.g.,  work of Paz and Sinha in \cite{decQC}). 
In adopting the influence functional scheme, one would be working with
 the density matrix of the universe, and the propagators of the reduced
 density matrix would be replacing the \$ matrix
of Hawking and Page \cite{Haw82,Page86} (similar in-out and in-in 
boundary condition difference  would matter). 
This would also offer a new angle towards the issues of
  unitarity and information loss  in quantum gravity.

\subsubsection{Tunnelling with Topology Change}  

Just as particle creation occurs when vacuum fluctuations of a quantum field
 get strong, pair creation of black holes may become important when metric
  fluctuations are large. One expects topology change in the spacetime to 
  occur at the Planck energy via tunneling. This is also part of the
   activities in a spacetime foam
 which has been studied by Hawking
  and his associates for a long time (see references in the spacetime foam section).
   In approaching these problems usually one defines the end states in terms of Lorentzian
geometry and describes the tunneling process by the Euclidean instanton method.
Finding the joining solution between two end states is not simple though, as it is not
so well defined. 

Similar to particle creation one may expect to cast the black hole pair creation
 as a dissipative process. If so, one would also need to work
in real time dynamics. The backreaction of these pair creation processes 
     is expected to be strong at the Planck energy. So the same set of issues
will arise as before. For example, how  would pair production 
of particles and black holes associated with topology change alter the tunneling rate and the
topology change itself? Our current understanding has not reached 
this level of  sophistication but these are important issues to think about.

\subsection{Nucleation of Black Holes from Curved Spacetime and
 Growth of Fluctuations and Forms}

From earlier discussions we see that vacuum instability and phase transition
may play an important role in revealing the structure of spacetime at the
Planck scale.  Ideally we wish to first formulate a quantum field theoretical
description of nucleation problem for first order phase transitions in general,
and then examine specific and related problems in gravity such as nucleation
of black holes from hot flat space \cite{GPY}, 
black hole pair creation in de Sitter universe \cite{BHpair}. The first problem
was studied by a number of authors in the 80's \cite{GPY,GinPer,YorWhi}
using Euclidean instanton methods to calculate the probability of nucleation.
If one could cast this problem in the form of a Langevin or Fokker-Planck equation 
we can reexamine this process as a dynamical critical phenomenon in real time.
Similarly we wish to carry out a first-principles quantum field theoretical
description of spinodal decomposition for the second order phase transitions.
This latter project we have just started with application to defect formation
in the early universe \cite{Greg}.

Advances in far-from-equilibrium
sciences in the last decade show that correlations and noise in
nonlinear systems are responsible for a great variety of structures
and forms \cite{Zia,KPZ}. Planck scale fluctuations can be the germinating
source for large scale structures in the universe. Noise-induced phase 
transition is an important class of problems originally studied by Kramer
for chemical kinetics. It is now applied with techniques from
stochastic gravity by Calzetta and Verdaguer \cite{CalVer} to the early
universe. They found the probability of a universe making such a phase 
transition to be very close to that of quantum tunneling studied earlier
by Vilenkin \cite{Vil} in the so-called `Birth of the Universe
from Nothing' scenario.  I only wish to add one observation. The proximity
of these two results appears to me not a plain accident. Let us  ponder on the relation
between noise-induced transition versus quantum tunneling.  While the former
usually refers to thermal noise (at finite temperature) in an environment, 
the latter refers to quantum noise (vacuum fluctuations). Even though one
does not stipulate an environment for quantum tunneling, quantum fluctuations
are ubiquitous and free (not quite: they are attached as the coarse-grained
leftovers from activities in the high energy sector, as reflected in some generalized
uncertainty principle). So the real difference is between thermal and 
vacuum fluctuations-induced effects. The relation between these fluctuations
in terms of their effect on decoherence has been studied before 
\cite{HuZha,AnaHal} in the context of finding an uncertainty relation at
finite temperature. Looking at the problem in another way, in terms of the
correlation hierarchy, quantum mechanical description invokes only the
lowest order correlations. At higher energy or with finer resolutions,
higer order correlations will partake more in the tunneling or transition
process. One can use the Schwinger-Dyson hierarchy and correlation noise
to put quantum and thermal fluctuations on the same footing. Neither of them 
need an environment nor a temperature stipulation. If we can relate these
two classes  of processes  (quantum and statistical mechanical) we
may find a way to deal with particle creation and tunneling together 
-- quantum or noise-induced -- in a unified real time formalism.


\subsection{Wave Propagation in Random Geometry and Simplicial Gravity}

In a recent paper Shiokawa and I  \cite{HuShi} studed some novel 
effects associated with electromagnetic wave
propagation in a Robertson-Walker universe and the Schwarzschild spacetime
with a small amount of metric stochasticity.
By showing the formal equivalence
of the wave equations in curved spacetimes with (flat space) wave propagation
in a material media and identifying the dependence of the refractive index
on the metric components, one can introduce metric fluctuations as a stochastic
component in the permittivity function and borrow the insights from known results 
of wave propagation in random media. We
find that localization of electromagnetic waves occurs
in a Robertson-Walker universe with time-independent metric stochasticity,
while time-dependent metric stochasticity
induces exponential instability in the particle production rate.
For the Schwarzschild metric, time-independent randomness
can decrease the total luminosity  of Hawking radiation
due to multiple scattering of waves
outside the black hole and
gives rise to event horizon fluctuations and thus
fluctuations in the Hawking temperature.

In their work the source of metric stochasticity is represented by a stochastic
component in the permitivity function. It is desirable to give a microscopic 
derivation of metric stochasticity.
Stochastic components in the metric can be induced by primordial gravitational waves,
topological defects in the sub-Planckian scale,
or intrisic metric fluctuations of background spacetimes at the Planck scale.
We should be able to calculate these components with the help of stochastic
gravity.
Their detection and analysis can provide valuable information about the state
of the early universe and black holes.
After this one can probe into wave propagation in random geometry itself
\cite{dyntriQG} via random potentials.
Eventually one should connect this to simplicial gravity \cite{simpQG}.
In addition to seeking the continuum limit from discrete geometries, it is
of interest to examine if possible disorder-order transition can arise
from stochastic spacetimes, and whether one could use this to divide the
effective (low energy, ordered or smoothed-out phase of)
spacetime into universality classes.

\subsection{Planck scale resonance states}

Following the progression from hydrodynamics to kinetic theory and
quantum micro-dynamics, one may ask if there could exist quasi-stable structures
at energy scales slightly higher than (or observation scales finer than)
the semiclassical scale. Assuming that string theory is the next level micro-theory,
does there exist quasi-stable structures between that and general relativity?
This is like the existence of resonance states (as quasi-stable particles)
beyond the stable compounds of quarks (baryons) or quark-antiquarks (mesons).
Viewed in the conceptual framework of kinetic theory, there could exist such
states, if the interparticle reaction times (collision and exchange)
and their characteristic dynamics (diffusion and dissipation) become
commensurate at some energy scale. (Turbulance in the nonlinear regime
could show up in these intermediate states). In the framework of decoherent
history discussed above, it could also provide metastable quasi-classical
structures. It would be interesting to find out if such structures can
in principle exist around the Planck scale. This question is stimulated
by the hydrodynamic viewpoint, but the resolution would probably have to
come from a combination of efforts from both the top-down and the
bottom-up approaches. Deductions from high energy string theories would also
benefit from knowing what different collective states are likely to exist
in the low energy physics of general relativity and semiclassical gravity.

\subsection{Spacetime Foams}

The beautiful, old and  alluring ideas of Wheeler \cite{Wheeler}
 on metric fluctuations  and spacetime foams have only  seen intermittant
  meaningful developements in the last thirty-five years since it was
   conceived, foremost by Hawking and his associates.
His work on quantum gravitational bubbles \cite{GravBub},   
wormholes and baby universes \cite{BabyWorm} ,
virtual black holes \cite{virBH} and black hole pair creation
\cite{BHpair} provided a solid base for such inquires. 
At the Planck scale geometric and topological fluctuations of spacetime
 are expected to be important. At a scale close to but larger than
 the Planck scale, stochastic gravity can provide a good physical depiction.
The extensively developed tools and concepts there can help one treat
the coarse-grained state of these  `building blocks' of spacetime foams and
come up with quantitative descriptions and predictions for low energy
 phenomenology. Metric fluctuations induced by quantum matter fields 
 (including gravitons) in the backreaction problems we have studied so
 far is perhaps the simplest and the most ubiquitous type of ingredients in 
 the spacetime foam.
We know them quantitatively by the noise or the correlation functions
 (see examples given at the beginning for weakly inhomogeneous cosmological 
spacetimes and far-field thermal black hole background).
The use of open system concepts enables one to view them as thermal baths \cite{Garay}
in the most naive approximation such as in the Fokker-Planck limit
 (Markovian behavior at high temperature  Ohmic bath in the case of bilinear
 coupling between the system and bath), but one lesson we learned
 from stochastic gravity is that these `noises' are by no means trivial,
as they contain precious information about the substructures and their 
constitution at a higher energy level. It would be interesting to examine 
the low energy remnants of the other types
of spacetime foams mentioned above. If we view them as an environment
 interacting with the classical geometry (which actually is the mean value
 taken with respect to all possible stochastic source distributions) 
 and study their behavior with the right model in
nonequilibrium statistical mechanics, one can get a rich physical picture 
 with quantitative informations (dissipation, diffusion, correlation, 
 decoherence). For example, virtual black holes can, according to a recent 
 suggestion \cite{Garay}, be representated at low energy by effective bilocal
 couplings. Hawking {\it et al} reasoned that spacetime are made up of three kinds
  of basic building blocks of topological classes: $S^2\times S^2, K3, CP^2$,
   and gravitational bubbles are believed to be their quantum fluctuations.
It is not easy to deal with these topological fluctuations, but in an
 effective description the vertex for the bubble scattering can be viewed
  as arising from the exchange of very large number of gravitons. 
  From this one can construct an open system model for multi-graviton
exchange and come up with a stochastic gravity version of this type 
of spacetime foam contribution.  Wormholes are more complicated as they
 are multiply-connected. One  can perform the same low-energy reduction even
  for   D-branes and talk about a D-foam background \cite{Ellis}.
  Even though these calculations cannot tell us the details of the basic 
  constituents of  spacetime but Planck scale spacetime fluctuations
   are a direct result of the activities of these substructures.
   Since they will affect all the
 physics happening at lower energies, they are worthy of much closer
 scrutiny. It is the only hope for us earthlings confined by the shackles 
of low energy  to fathom the blue yonder. \\

{\bf Acknowledgements}  This summary is based on work done over the past 5 years with my
students, associates and colleagues. I thank Esteban Calzetta, Antonio Campos,  Ted Jacobson, 
Phillip Johnson, Andrew Matacz, Diego Mazzitelli, Juan Pablo Paz, Nicholas Phillips,
Stephen Ramsey, Alpan Raval,  Kazutomu Shiokawa, Sukanya Sinha, Greg Stephens and 
Yuhong Zhang for stimulating discussions and fruitful collaboration. Any unjustified claim, 
or false speculation, is of course solely my responsibility. I thank Enric Verdaguer for 
providing me a copy of his talk and a summary of his nice work with 
Rosario Mart\'in. Esteban and Enric also took the trouble to read through 
my draft and gave useful comments.   I also enjoyed listening to a recent 
seminar of Professor G. Volovik, 
who seems to share a similar view on many basic issues. For the conference, 
special appreciation  goes to the organizers of this meeting, especially Edgar Gunzig, 
who not only provided us with  the warm hospitality and congenial atmosphere for 
discussions, but also flew in the chefs and ordered the beautiful scenary. 
This work is supported in part by NSF PHYS 98-00967.


\begin{references}



\bibitem{BirDav}
N. D. Birrell, P. C. Davis, {\it Quantum Fields in Curved Space}
 (Cambridge University Press, New York, 1982).

\bibitem{cpcbkr} 
Ya. Zel'dovich and A. Starobinsky, Zh. Eksp. Teor. Fiz {\bf 61}, 2161 (1971)
[Sov. Phys.- JETP {\bf 34}, 1159 (1971)]
L. Grishchuk, Ann. N. Y. Acad. Sci. 302, 439 (1976).
B. L. Hu and L. Parker, Phys. Lett. 63A, 217 (1977).
B. L. Hu  and L. Parker, Phys. Rev. {\bf D17}, 933 (1978).
F. V. Fischetti, J. B. Hartle and B. L. Hu, Phys. Rev. {\bf D20}, 1757 (1979).
J. B. Hartle and B. L. Hu, Phys. Rev. {\bf D20}, 1772 (1979).
{\bf 21}, 2756 (1980).
J. B. Hartle, Phys. Rev. D23, 2121 (1981).
P. A. Anderson, Phys. Rev. D28, 271 (1983); D29, 615 (1984).

\bibitem{CH87}
E. Calzetta and B. L. Hu, Phys. Rev. {\bf D35}, 495 (1987)

\bibitem{ctp}
J. Schwinger, J. Math. Phys. {\bf 2} (1961) 407;
P. M. Bakshi and K. T. Mahanthappa,
J. Math. Phys. 4, 1 (1963), 4, 12 (1963);
L. V. Keldysh, Zh. Eksp. Teor. Fiz. {\bf 47 }, 1515 (1964)
[Engl. trans. Sov. Phys. JEPT {\bf 20}, 1018 (1965)].
G. Zhou, Z. Su, B. Hao and L. Yu, Phys. Rep. {\bf 118}, 1 (1985);
Z. Su, L. Y. Chen, X. Yu and K. Chou, Phys. Rev. {\bf B37}, 9810
(1988).
B. S. DeWitt, in {\it Quantum Concepts in Space and Time},
ed. R. Penrose and C. J. Isham (Claredon Press, Oxford, 1986);
R. D. Jordan, Phys. Rev. D33, 44 (1986).
E. Calzetta and B. L. Hu, Phys. Rev. {\bf D35}, 495 (1987).
A. Campos and E. Verdaguer, Phys. Rev. {\bf D49}, 1861 (1994).

\bibitem{Physica} 
  B.-L. Hu, Physica {\bf A158}, 399 (1989).

\bibitem{decQC} 
C. Kiefer, Clas. Quant. Grav. {\bf 4}, 1369 (1987);
 J. J. Halliwell, Phys. Rev. {\bf D39}, 2912 (1989); T. Padmanabhan, {\it ibid.} 2924 (1989).
 B. L. Hu ``Quantum and Statistical Effects in Superspace Cosmology'' in 
{\it Quantum Mechanics in Curved Space
time}, ed. J. Audretsch and V. de Sabbata (Plenum, London 1990)
E. Calzetta, Class. Quan. Grav. {\bf 6}, L227 (1989); Phys. Rev. 
{\bf D 43}, 2498 (1991); J. P. Paz and S. Sinha, Phys. Rev. {\bf D44}, 1038 (1991); 
{\it ibid} {\bf D45}, 2823 (1992).
 B. L. Hu, J. P. Paz and S. Sinha,
''Minisuperspace as a Quantum Open System'' in {\it Directions in General Relativity} 
Vol. 1, (Misner Festschrift) eds B. L. Hu , M. P. Ryan and C. V. Vishveswara
 (Cambridge Univ., Cambridge, 1993)

\bibitem{qos}
See, e.g., E. B. Davies, {\it The Quantum Theory of Open
Systems} (Academic Press, London, 1976); K. Lindenberg and B. J. West,
{\it The Nonequilibrium Statistical Mechanics of Open and Closed Systems}
(VCH Press, New York, 1990); U. Weiss, {\it Quantum Dissipative Systems}
(World Scientific, Singapore, 1993) 

\bibitem{if}
R. Feynman and F. Vernon, Ann. Phys. (NY) {\bf 24}, 118 (1963).
R. Feynman and A. Hibbs, {\it Quantum Mechanics and Path Integrals},
(McGraw - Hill, New York, 1965).
A. O. Caldeira and A. J. Leggett, Physica {\bf 121A}, 587 (1983).
H. Grabert, P. Schramm and G. L. Ingold, Phys. Rep. {\bf 168}, 115
(1988).
B. L. Hu, J. P. Paz and Y. Zhang, Phys. Rev. {\bf D45}, 2843 (1992);
{\bf D47}, 1576 (1993)

\bibitem{qbm}
A. O. Caldeira and A. J. Leggett, Physica {\bf 121A}, 587 (1983);
Ann. Phys. (NY) {\bf 149}, 374 (1983).
H. Grabert, P. Schramm and G. L. Ingold, Phys. Rep. {\bf 168}, 115 (1988).
B. L. Hu, J. P. Paz and Y. Zhang, Phys. Rev. {\bf D45}, 2843 (1992);
{\bf D47}, 1576 (1993) 

\bibitem{Tsukuba}
B.. L. Hu, ``Statistical Mechanics and Quantum Cosmology'',
{\it Proc. Second International Workshop on Thermal Fields
and Their Applications}, eds. H. Ezawa et al (North-Holland, Amsterdam, 1991).

\bibitem{envdec}
W. H. Zurek, Phys. Rev. D24, 1516 (1981); D26, 1862
(1982); in {\it Frontiers of Nonequilibrium Statistical Physics}, 
ed. G. T. Moore and M. O. Scully (Plenum, N. Y., 1986); Physics Today {\bf 44}, 36 (1991);
 E. Joos and H. D. Zeh, Z. Phys. B59, 223 (1985); 
A. O. Caldeira and A. J. Leggett, Phys. Rev. {\bf A 31}, 1059 (1985);
W. G. Unruh and W. H. Zurek, Phys. Rev. D40, 1071 (1989);  
B. L. Hu, J. P. Paz and Y. Zhang, Phys. Rev. {\bf D45}, 2843 (1992);
W. H. Zurek, Prog. Theor. Phys. 89, 281 (1993).   D. Giulini 
{\it et al}, {\it Decoherence and the Appearance of a Classical World in Quantum Theory} 
(Springer Verlag, Berlin, 1996)

\bibitem{conhis}
R. B. Griffiths, J. Stat. Phys. {\bf 36}, 219 (1984).
R. Omn\'es, J. Stat Phys. {\bf 53}, 893, 933, 957 (1988); Ann. Phys. (NY)
{\bf 201}, 354 (1990); Rev. Mod. Phys. {\bf 64}, 339 (1992); {\it The
Interpretation of Quantum Mechanics} (Princeton UP, Princeton, 1994).
M. Gell-Mann and J. B. Hartle, in {\it Complexity, Entropy and the Physics
of Information}, ed. by W. H. Zurek (Addison-Wesley, Reading, 1990);
Phys. Rev. {\bf D47}, 3345 (1993) ;
J. B. Hartle, ``Quantum Mechanics of Closed
Systems'' in {\it Directions in General Relativity} Vol. 1, eds B. L. Hu,
M. P. Ryan and C. V. Vishveswara (Cambridge Univ., Cambridge, 1993).
H. F. Dowker and J. J. Halliwell, Phys. Rev. {\bf D46}, 1580 (1992);
T. Brun, Phys. Rev. {\bf D47}, 3383 (1993);
J. P. Paz and W. H. Zurek, Phys. Rev. {\bf D48} 2728 (1993);
J. Twamley, Phys. Rev. {\bf D48}, 5730 (1993).
F. Dowker and A. Kent, Phys. Rev. Lett {\bf 75, } 3038
(1995); J. Stat. Phys. {\bf 82, }1575 (1996); 
A. Kent, Phys. Rev. {\bf A54,} 4670 (1996); 
Phys. Rev. Lett {\bf 78, } 2874 (1997); Phys. Rev. Lett {\bf 81, } 1982 (1998).

\bibitem{FDR}
    W. Bernard and H. B. Callen,
          Rev. Mod. Phys. {\bf 31}, 1017 (1959);
    R. Kubo,
          Rep. Prog. Phys. {\bf 29}, 255 (1966);
    L. Landau, E. Lifshitz and L. Pitaevsky,
          {\sl Statistical Physics},
          (Pergamon, London, 1980), Vol. 1;
    R. Kubo, M. Toda and N. Hashitsume,
          {\sl Statistical Physics II},
          (Springer-Verlag, Berlin, 1985).

\bibitem{LRT} 
See, e.g., R. Kubo, M. Toda and N. Hashitsume, {\it Statistical Physics}
 Vol 2, (Springer-Verlag, Berlin, 1991) Chapter 4.

\bibitem{GelHar2}
M. Gell-Mann and J. B. Hartle, Phys. Rev. {\bf D47}, 3345 (1993) 

\bibitem{noisefield}
Y. Zhang, Ph. D. Thesis, University of Maryland (1991).
B. L. Hu, J. P. Paz and Y. Zhang, ``Quantum Origin of
Noise and Fluctuation in Cosmology'' in {\it The Origin of Structure in the Universe}
 Conference at Chateau du Pont d'Oye, Belgium, April, 1992, 
ed. E. Gunzig and P. Nardone (NATO ASI Series) (Plenum Press, New York, 1993) p. 227.

\bibitem{HM2}
B. L. Hu and A. Matacz, Phys. Rev. D49, 6612 (1994).

\bibitem{CH94}
E. Calzetta and B. L. Hu, Phys. Rev {\bf D49}, 6636 (1994).

\bibitem{Su}
Z. Su, L. Y. Chen, X. Yu and K. Chou, Phys. Rev. {\bf B37}, 9810
(1988).

\bibitem{ifdf}
H. F. Dowker and J. J. Halliwell, Phys. Rev. {\bf D46}, 1580 (1992);
J. P. Paz and W. H. Zurek, Phys. Rev. {\bf D48} 2728 (1993)

\bibitem{dch}  E. Calzetta and B. L. Hu, ``Decoherence of Correlation
Histories'' in {\it Directions in General Relativity, Vol II: Brill
Festschrift}, eds B. L. Hu and T. A. Jacobson (Cambridge University Press, Cambridge, 1993).

\bibitem{CV94}
A. Campos and E. Verdaguer, Phys. Rev. {\bf D49}, 1861 (1994).

\bibitem{MarVer}
R. Martin and E. Verdaguer, Int. J. Theor. Phys. (1999 this issue). gr-qc/9812063, 9811070.

\bibitem{Banff}
B. L. Hu, in {\it Proceedings of the Third International Workshop on Thermal Fields and
its Applications}, CNRS Summer Institute, Banff,
August 1993, edited by R. Kobes and G. Kunstatter (World Scientific, Singapore, 1994).

\bibitem{HuSin}
B. L. Hu and S. Sinha,
{\bf D 51}, 1587 (1995).

\bibitem{HM3}
B. L. Hu and A. Matacz, Phys. Rev. {\bf D51}, 1577 (1995).

\bibitem{CV96}
A. Campos and E. Verdaguer, Phys. Rev. {\bf D53}, 1927 (1996).

\bibitem{FulPar}
S. A.  Fulling and L. Parker, Ann. Phys. (N. Y. )  87, 176  (1974)

\bibitem{CCV}
    E. Calzetta, A. Campos and E. Verdaguer,
          Phys. Rev. D {\bf 56}, 2163 (1997).

\bibitem{Hor}
G. Horowitz,  Phys. Rev. {\bf D21}, 1445 1980)

\bibitem{LomMaz}
F. C. Lombardo and F. D. Mazzitelli, Phys. Rev. {\bf 55}, 3889 (1997)

\bibitem{LMR}
F. C. Lombardo and F. D. Mazzitelli, Phys. Rev. {\bf 58}, 024009 (1998).
F. C. Lombardo and F. D. Mazzitelli and J. Russo, Phys. Rev. D (1999).

\bibitem{JohHu}
P. Johnson and B. L. Hu,  ``Quantum Stochastic Theory of Relativistic 
Particle- Field Interaction:  Effect of Strong Fields on the Trajectory of a  Particle Detector.
in preparation.

\bibitem{cddn}
E. Calzetta and B. L. Hu, ``Correlations, Decoherence,
Disspation and Noise in Quantum Field Theory'', in {\it Heat Kernel
Techniques and Quantum Gravity}, 
ed. S. Fulling (Texas A\& M Press, College Station 1995). hep-th/9501040


\bibitem{StoBol}
E. Calzetta and B. L. Hu, ``Nonequilibrium Quantum Fields: Master Effective Action,
 Correlation Noise and Stochastic Boltzmann Equation"
(in preparation)

\bibitem{Vishu}
B. L. Hu, Alpan Raval and S. Sinha, ``Notes on Black Hole Fluctuations and Backreaction, in 
{\it Black Holes, Gravitational Radiation and the Universe } eds. B. R.  Iyer and  
B. Bhawal (Kluwer Academic Publishers,  Dordtrecht, 1999)

\bibitem{Phillips} 
N. G. Phillips, Ph. D. Thesis, University of Maryland (1999)

\bibitem{HPR}
B. L. Hu, N. Phillips and A. Raval, {\it Fluctuations of the Energy Mementum 
Tensor of a Quantum Field in a Black Hole Spacetime} (in preparation)

\bibitem{HRS}
B. L. Hu, Alpan Raval and S. Sinha, {\it Backreaction of a Radiating 
Quantum Black Hole and Fluctuation-Dissipation Relation} (in preparation)

\bibitem{CalVer}
E. Calzetta and E. Verdaguer, Phys. Rev. D59 (1999)

\bibitem{CamHu} 
A. Campos and B. L. Hu, Phys. Rev. D58,  125021 (1998)

\bibitem{Wheeler}
J. A. Wheeler,  Ann. Phys. (N. Y.) 2, 604 (1957); 
{\it Geometrodynamics} (Academic Press, London, 1962);
in {Relativity, Groups and Topology}, eds B. and C. DeWitt
(Gordon and Breach, New York, 1964).

\bibitem{Ford}
L. H. Ford, Phys. Rev {\bf D51}, 1692 (1995). 

\bibitem{KuoFor}
C.-I. Kuo and L. H. Ford, Phys. Rev. D47, 4510 (1993)

\bibitem{PhiHu}
N. Phillips and  B. L. Hu, Phys. Rev. D {\bf 55}, 6123  (1997)

\bibitem{CH95}
    E. Calzetta and B.-L. Hu,
          Phys. Rev. D {\bf 52}, 6770 (1995);
    E. Calzetta and S. Gonorazky,
          Phys. Rev. D {\bf 55}, 1812 (1997);
    A. Matacz,
          Phys. Rev. D {\bf 55}, 1860 (1997);
          {\bf 56}, R1836 (1997).

\bibitem{ChrDeW}
 B. S. DeWitt,  {\it Dynamical Theory of Groups and Fields} 
 (Gordon and Breach, 1965);  Phys. Rep. {\bf C19}, 295 (1975).
S. Christensen, Phys. Rev. D14, 2490  (1976)

\bibitem{AndHis}
P. R. Anderson, W. A. Hiscock and D. A. Samuel, Phys. Rev. Lett. 70,
1739 (1993); Phys. Rev. D51, 4337 (1995)

\bibitem{Page}
D. N. Page, Phys. Rev. D25, 1499 (1982).

\bibitem{Ford97}
L. H. Ford,
{\it Cosmological and Black Hole Horizon Fluctuations},  gr-qc/9704050.
\bibitem{CEIMP} A. Casher, F. Englert, N. Itzhaki, and R. Parentani,
{\it Black Hole Horizon Fluctuations}, hep-th/9606106.

\bibitem{Sorkin}
 R.D. Sorkin,
{\it How Wrinkled is the Surface of a Black Hole?}, gr-qc/9701056.

\bibitem{BFP}
C. Barrab\`es, V. Frolov and R. Parentani, 
 ``Metric Fluctution Correction to Hawking Radiation" gr-qc/9812076

\bibitem{bhbkrn}
P. Anderson {\it et al}, in {\it Heat Kernel Techniques and Quantum Gravity},
Vol. 4 of {\it Discourses in Mathematics and Its Applications},
Winnipeg, 1994, edited by S. A. Fulling (Texas A\&M University Press,
College Station, TX, 1995)

\bibitem{York}
J. W. York, Jr., Phys. Rev. D28, 2929 (1983); D31, 775 (1985); D33,
2092 (1986)

\bibitem{Bardeen}
J. M. Bardeen, Phys. Rev. Lett. 46, 382 (1981).
P. Hajicek and W. Israel, Phys. Lett. 80A, 9 (1980)

\bibitem{Masser} 
S. Massar, Phys. Rev. {\bf D 52}, 5857 (1995).

\bibitem{CanSci}
P. Candelas and D. W. Sciama, Phys. Rev. Lett. {\bf 38}, 1372 (1977)

\bibitem{Mot}
E. Mottola,  Phys. Rev. D33, 2136 (1986)

\bibitem{Bari} 
A. Campos and B. L. Hu, Int. J. Theor. Physics (April 1999)

\bibitem{GPY}
D. J. Gross, M. J. Perry and L. G. Yaffe,
          Phys. Rev. D {\bf 25}, 330 (1982).

\bibitem{Reb}
A. Rebhan,
          Nucl. Phys. {\bf B351}, 706 (1991).

\bibitem{ABFT}
A. P. de Almeida, F. T. Brandt and J. Frenkel,
          Phys. Rev. D {\bf 49}, 4196 (1994),
              F. T. Brandt and J. Frenkel, Phys. Rev. D58, (1998)
              and references therein.

\bibitem{meso} 
B. L. Hu, in {\it Quantum Classical  Correspondence} eds. D. S. Feng and B. L. Hu
 (Interntional Press, Boston, 1997)

\bibitem{Sak}
A. D. Sakharov, ``Vacuum Quantum Fluctuations in Curved Space and the Theory of
Gravitation'' Doklady Akad. Nauk S. S. R. 177, 70-71 (1987)
[Sov. Phys. - Doklady 12, 1040-1041 (1968)]. See also S. L. Adler, Rev. Mod.
Phys. 54, 729 (1982)

\bibitem {grhydro}
B. L. Hu, ``General Relativity as Geometro-Hydrodynamics"   gr-qc/9607070

\bibitem{CH97}
 E. Calzetta and B. L. Hu, Phys. Rev. {\bf D55}, 1795 (1997).

\bibitem{mesobooks}
See, e.g., B. L. Altshuler, P. A. Lee and R. A. Webb, eds,
     {\it Mesoscopic Phenomena in Solids} (North Holland, Amsterdam, 1991).
B. K. Kramer, ed., {\it Quantum Coherence in Mesoscopic Systems}
                   (Plenum Press, New York, 1991).
W. P. Kirk and M. A. Reed, eds, {\it Nanostructures and Mesocopic Systems}
                           (Academic Press, San Diego, 1992).
Y. Imry, {\it Introduction to Mesoscopic Physics} (J. Wiley, N. Y. 1997)

\bibitem{CalLeg83}
A. O. Caldeira and A. J. Leggett, Ann. Phys. (N. Y.) 149, 374 (1993).

\bibitem{Weinberg}
S. Weinberg, Quantum Field Theory Vol 1, 2  (J. Wiley, New York, 1996)

\bibitem{cgea}
B. L. Hu and Y. Zhang, ``Coarse-Graining, Scaling, and Inflation"
Univ. Maryland Preprint 90-186 (1990);
B. L. Hu, in {\it Relativity and Gravitation: Classical
and Quantum} Proc. SILARG VII, Cocoyoc, Mexico 1990.
eds. J. C. D' Olivo et al (World Scientific, Singapore 1991).

\bibitem{JabVol}
T. Jacobson, ``Introduction to Black Hole Microscopy"  hep-th/9510026

\bibitem{JacUnr}
W. G. Unruh, Phys. Rev. Lett. 46, 1351 (1981); Phys. Rev. D51, 2827
(1995)
T. Jacobsen, Phys. Rev. D44, 1731 (1991); Phys. Rev. D53, 7082 (1994)

\bibitem{Spain}
B. L. Hu, ``Fluctuation, Dissipation and Irreversibility
in Cosmology'' in {\it The Physical Origin of Time-Asymmetry}, Huelva,
Spain, 1991 eds. J. J. Halliwell, J. Perez-Mercader and W. H. Zurek
(Cambridge University Press, Cambridge, 1994).

\bibitem{dechyd}  J. B. Hartle, R. Laflamme and D. Marolf, Phys. Rev. D51,
7007 (1995); T. Brun and J. J. Halliwell, Phys. Rev. {\bf D54}, 2899 (1996);
J. J. Halliwell, Phys. Rev. {\bf D58}, 105015 (1998); C. Anastopoulos,
gr-qc/9805074; T. Brun and J. B. Hartle, quant-phys/9808024.

\bibitem{WheelerBianchi}.
See, e.g., C. W.  Misner, K. S.  Thorne and J. A. Wheeler, {\it Gravitation}
 (Freeman, S. F. 1971) Chapter 15

\bibitem{string}
M. B. Green, J. H. Schwarz and E. Witten,
{\it Superstring Theory} (Cambridge University, Cambridge, 1990).
E. Witten, Physics Today 49, 24 (1996). J. Polchinsky, {\it  Superstring Theory }
(Cambridge University Press, Cambridge 1998).

\bibitem{loop}
A. Ashtekar, {\it Lectures on Non-Perturbative Canonical Gravity}
(World Scientific, Singapore, 1991); {\it Knot Theory and Quantum Gravity}
ed. J. Baez (Oxford University, London, 1995). 
R. Gambini and J. Pullin, {\it Loops, Knots, Gauge Theories and Quantum Gravity}
(Cambridge University Press, Cambridge, 1996).
A. Ashtekar, K. Baez, A. Corichi and  K. Krasnov, Phys. Rev. Lett. 80, 904 (1998).

\bibitem{simpQG}
T. Regge, Nuovo Cimento 19, 558 (1961).
For  recent work, see, e.g.,
J. B. Hartle, J. Math. Phys. 26, 804 (1985); 27, 287 (1986); 30, 452 (1989).
H. W. Hamber, in {\it Critical Phenomena, Random Systems, Gauge Theories}
      1984 Les Houches Summer School, eds K. Osterwalder and R. Stora
      (North Holland, Amsterdam, 1986).
H. W. Hamber, Nucl. Phys. B (Proc. Suppl) 20, 728 (1991); 25A, 150 (1992);
     Phys. Rev. D45, 507 (1992); Nucl. Phys. B400, 347 (1993);
R. M. Williams and P. A. Tucker, Class. Quant. Grav. 9, 1409 (1992).
H. W. Hamber and R. M. Williams, Phys. Rev. D47, 510 (1993);
      Nucl. Phys. 415, 463 (1994)

\bibitem {BekHaw}
J. D. Bekenstein, Phys. Rev. D7, 1333 (1973).
S. W. Hawking, Commun. Math. Phys. {\bf 43}, 199 (1975).

\bibitem {tdbhent}
R. M. Wald, Phys. Rev. D48, R3427 (1993).
T. Jacobson, G. Kang, and R. Myers, Phys. Rev. D49, 6587 (1994).
T. Jacobson, ``Black Hole Entropy and Induced Gravity" gr-qc/9404039.
M. Banados, C. Teitelboim and J. Zanelli, Phys. Rev. Lett. 72, 957 (1994).
J. D. Bekenstein, `` Do We Understand Black Hole Entropy?"
 Proc. Seventh Marcel Grossmann Meeting, Stanford University, 1994,  gr-qc/9409015.
D. M. Page, ``Black Hole Information", in {\it Proc. 5th Canadian Conference
on General Relativity and Relativistic Astrophysics} 
eds. R. B. Mann and R. G. McLenaghan (World Scientific, Singapore, 1994) hep-th/9305040.
V. P. Frolov, D. V. Fursaev and A. I. Zelnikov, ``Black Hole Entropy:
Off-Shell vs On-Shell" hep-th/9512184; ``Black  Hole Entropy and Induced Gravity"
hep-th/9607104

\bibitem{JacEqState}
T. Jacobson, Phys. Rev. Lett. 75, 1260 (1995)

\bibitem {smbhent}
L. Susskind and J. Uglam, Phys. Rev. D50, 2700 (1994); 
D. Kabat, S. H. Shenker, and M. J. Strassler, Phys. Rev. D52, 7027 (1995)
J. D. Bekenstein and V. F. Mukhanov, Phys. Lett. B 360, 7 (1995).

\bibitem{strbhent}
A. Strominger and C. Vafa, Phys. Lett. B379, 99 (1996)
G. T. Horowitz, ``The Origin of Black Hole Entropy in String Theory"
in the Proceedings of the Pacific Conference on Gravitation and Cosmology,
Seoul, Korea, Feb, 1996. gr-qc/9604051
G. Horowitz and J. Polchinski, Phys. Rev. D55, 6189 (1997)
J. M. Maldacena, ``Black Holes and D-Branes" Nucl. Phys. Proc. Suppl. 
{\bf 61A} 111 (1998)

\bibitem {AshErice}
A. Ashtekhar, in {\it String Gravity and Physics at the Planck Energy Scale}
eds. N. Sanchez and A. Zichichi (Kluwer, Dordrecht, 1996)

\bibitem{simpQG}
T. Regge, Nuovo Cimento 19, 558 (1961).
J. B. Hartle, J. Math. Phys. 26, 804 (1985); 27, 287 (1986); 30, 452 (1989).
H. W. Hamber, in {\it Critical Phenomena, Random Systems, Gauge Theories}
      1984 Les Houches Summer School, eds K. Osterwalder and R. Stora
      (North Holland, Amsterdam, 1986).

\bibitem {PonReg}
G. Ponzano and T. Regge, "Semiclassical limit of Racah coefficients"
in {\it Spectroscopic and Group Theoretical Methods in Physics}
ed. F. Bloch (North Holland, Amsterdam, 1968)
J. Iwasaki, ``A reformulation of the Ponzano- Regge quantum gravity models
in terms of surfaces" (1994)
J. W. Barrett and T. J. Foxon, Class. Quant. Grav. 11, 543 (1994).
J. W. Barrett, ``Quantum gravity as topological quantum field theory" (1995)

\bibitem {dyntriQG}
For recent reviews, see, e.g., {\it Statistical Mechanics of Membranes
and Surfaces} eds D. Nelson et al (World Scientific, Singapore, 1989).
{\it Two-Dimensional Quantum Gravity and Random Surfaces} eds D. J. Gross,
T. Piran and S. Weinberg (World Scientific, Singapore, 1992).
{\it Fluctuating Geometries in Statistical Mechanics and Field Theory}, 
eds F. David, P. Ginsparg and J. Jinn-Justin,
Les Houches Session LXII 1994 (North-Holland, Amsterdam, 1996).
J. Ambj\/orn, M. Carfora and A. Marzuoli, {\it The Geometry of Dynamical 
Triangulations} (Springer, Berlin, 1997)


\bibitem{HK}
B. L. Hu, ``Cosmology as `Condensed Matter' Physics'' in Proc. Third  Asia
Pacific Physics Conference,
ed. Y. W. Chan et al (World Scientific, Singapore, 1988) Vol. 1, p. 301.
gr-qc/9511076

\bibitem {Sus}
L. Susskind, J. Math. Phys. 36, 6377 (1995)

\bibitem {tHo}
G. 't Hooft,  ``Quantization of Point Particles in 2+1 Dimensional Gravity
and Spacetime Discreteness" gr-qc/9601014

\bibitem{eft}  T. Appelquist and J. Carazzone, Phys. Rev. {\bf D11}, 2856
(1975) S. Weinberg, Phys. Lett. {\bf 83B}, 339 (1979); B. Ovrut and H. J.
Schnitzer, Phys. Rev. {\bf D21}, 3369 (1980); {\bf D22}, 2518 (1980) L.
Hall, {\sl Nucl. Phys. }{\bf B178}, 75 (1981). P. Lapage, in {\it From
Actions to Answers}, Proc. 1989 Adv. Inst. in Elementary Particle Physics,
eds T. DeGrand and D. Toussaint (World Scientific, Singapore, 1990) p. 483.
S. Weinberg, {\it The Quantum Theory of Fields} (Vol. 1)
(Cambridge University Press, Cambridge, 1995).

\bibitem{HarHu}
J. B. Hartle and B. L. Hu, Phys. Rev. {\bf D20}, 1772 (1979).
{\bf 21}, 2756 (1980).

\bibitem{Sch}
J. Schwinger, Phys. Rev. 82, 664 (1951).

\bibitem{Mot}
E. Mottola,  Phys. Rev. D33, 2136 (1986)

\bibitem{CH88}
E. Calzetta and B.-L. Hu,   Phys. Rev. D  {\bf 37}, 2878 (1988).

\bibitem{kft80}
 P. Danielewicz, Ann. Phys. (NY) 152, 239 (1984)
S-P. Li and L. McLerran, Nucl. Phys. B214, 417 (1983).

\bibitem{qustme}
Kadanoff and Baym, {\it Quantum Statistical Mechanics}
(Benjamin, New York, 1962).
D. F. Du Bois, in {\it Lectures in Theoretical Physics} Vol 9c,
edited by W. Brittin (Gordon and Breach, New York, 1967).
 B. Bezzerides and D. F. Du Bois, Ann. Phys. (NY) 70, 10 (1972).

\bibitem{kft90}
St. Mrowczynski and P. Danielewicz, Nucl. Phys. B342, 345 (1990).
St. Mrowczynski and U. Heinz, Ann. Phys. 229,1 (1994).
P. Henning, Phys Rep 253, 235 (1995);  Nucl Phys A582, 633 (1995).
 D. Boyanovsky, I. Lawrie and D-S. Lee, Phys. Rev. D54, 4013 (1996)
 E. Wang and U. Heinz, Phys. Rev. D53, 899 (1996)

\bibitem{KacLog}
M. Kac and J. Logan, ``Fluctuations'', in {\it Fluctuation Phenomena}
 edited by E. W. Montroll and J. L. Lebowitz (Elsevier, New
York, 1979), p.1

\bibitem{effaction}
R. Jackiw, Phys. Rev. {\bf D9}, 1686 (1974); J.
Iliopoulos, C. Itzykson and A. Martin, Rev. Mod. Phys. {\bf 47}, 165 (1975).

\bibitem{pi2}  H. D. Dahmen and G. Jona - Lasinio, Nuovo Cimento {\bf 52A},
807 (1962); C. de Dominicis and P. Martin, J. Math. Phys. {\bf 5}, 14
(1964); J. M. Cornwall, R. Jackiw and E. Tomboulis, Phys. Rev. {\bf D10},
2428 (1974); R. E. Norton and J. M. Cornwall, Ann. Phys. (NY) {\bf 91}, 106
(1975).

\bibitem{HarHor}
J. B. Hartle and G. Horowitz, Phys. Rev. D 24, 257 (1981)

\bibitem{KME}
Y. Kluger, E. Mottola and J. M. Eisenberg, Phys. Rev. D58, 125015 (1998)

\bibitem{mesotrans}
F. A. Buot, Phys. Rep. 234, 73-174 (1993).
S. Datta, {\it Electronic Transport in Mesoscopic Systems}
          (Cambridge University Press, Cambridge, 1995).
P. Sheng, {\it Introduction to Wave Scattering, Localization and Mesoscopic
           Phenomena} (Academic Press, New York, 1995).

\bibitem{gravscatt}
S. Chandrasekhar  and B. C. Xanthopoulos, Proc. Roy.  Soc. A408, 175 (1986);
 A410, 311 (1987)  
A. G. Mirzbekian and G. Vilkovisky,  gr-qc/9803006

\bibitem{Greg}
G. Stephens, E. Calzetta, B. L. Hu and S. A. Ramsey, Phys. Rev. D59,  045009 (1999)

\bibitem {Rub}
V. A. Rubakov, Nucl. Phys. B245, 481 (1991)

\bibitem {VacVil}
T. Vachaspati and A. Vilenkin, Phys. Rev. D37, 898 (1988)

\bibitem{RH1}
S. A. Ramsey and B. L. Hu, Phys. Rev. D56, 678  (1997)

\bibitem{CalLeg}
A. O. Caldeira and A. J. Leggett, Ann. Phys. (N. Y.) {\bf 149}, 374 (1983).

\bibitem{Vil}
A. Vilenkin, Phys. Rev. D27, 2848 (1983), D30, 509 (1984);
Phys. Lett. 117B, 25 (1985); Nucl. Phys. {\bf B226}, 527 (1983)

\bibitem {Haw82}
S. W. Hawking, Comm. Math. Phys. 87, 395 (1982)

\bibitem {Page86}
D. N. Page, Phys. Rev. D34, 2267 (1986)

\bibitem{GinPer}
P. Ginzparg and M. J. Perry, Nucl. Phys. B222, 245 (1983)

\bibitem{YorWhi}
B. F. Whiting and J. W. York, Jr.  Phys. Rev. Lett. 61, 1336 (1988)

\bibitem{Zia}
R. K. P. Zia, ``Driven Diffusive Systems" in
{\it Phase Transitions and Critical Phenomena}
Vol. 20, eds. C. Domb and J. Lebowitz (Academic Press, New York, 1995)

\bibitem{KPZ}
M. Kadar, G. Parisi and Y. C. Zhang, Phys. Rev. Lett. 56, 342 (1986)

\bibitem{HuZha}
B. L. Hu and Y. Zhang, Mod. Phys. Lett. A8, 3575 (1993);
 Int. J. Mod. Phys. 10, 4537 (1995)

\bibitem{AnaHal}
A. Anderson and J. J. Halliwell, Phys. Rev. D48, 2753 (1993).
A. Anastopoulos and J. J.  Halliwell, Phys. Rev. D51, 6870 (1995).
A. Anastopoulos, Phys. Rev. E53, 4711 (1996).

\bibitem{HuShi}
    B.-L. Hu and K. Shiokawa,
          Phys. Rev. D {\bf 57}, 3474 (1998).

\bibitem{GravBub}
S. W. Hawking, Nucl. Phys. B144, 349 (1978).
S. W. Hawking, D. N. Page and C. N. Pope, Nucl. Phys. B170 [FS1] 283 (1980).
N. P.  Warner,  Comm. Math. Phys. 86, 419 (1982).

\bibitem{BabyWorm}
S. W. Hawking, Phys. Rev. D37, 904 (1988).
S. Coleman, Nucl. Phys. B307, 867 (1988).
S. Coleman, J. B. Hartle, T. Piran and S. Weinberg, eds.
{\it Quantum Cosmology and Baby Universes} (World Scientific, Singapore, 1991)

\bibitem{virBH}
S. W. Hawking, Phys. Rev. D53, 3099 (1996).

\bibitem{BHpair}
D. Garfunkle, S. B. Giddings and A. Stominger, Phys. Rev. D 49, 958 (1994).
F. Dowker et al, Phys. Rev. D50, 2662  (1994).
S. W. Hawking, G. T. Horowitz and S. F. Ross, Phys. Rev. D51, 4302 (1995.)
R. Busso and S. W. Hawking,  Phys. Rev. 52, 5659 (1995) 54, 6312 (1996).

\bibitem{Garay}
S. Carlip, Phys. Rev. Lett. 79, 4071 (1997); Class. Quan. Grav. 15, 2629 (1998)
L. J. Garay, Phys. Rev. Lett. 80, 2508 (1998); Phys. Rev. D58, 124015 (1998)

\bibitem{Ellis}
J. Ellis, N. E. Mavromatos and D. V. Nanopoulos,``Quantum Decoherence in a D-Foam
Background"  Mod. Phys. Lett A. (1997)

\bibitem{Huang}
K. Huang, {\it Statistical Mechanics}, 2nd Ed. (John Wiley, New York, 1987).

\bibitem{Akhiezer}
A. I. Akhiezer and S. V. Peletminsky, {\it Methods of
Statistical Physics} (Pergamon, London, 1981).

\end{references}
\end{document}